\PassOptionsToPackage{unicode}{hyperref}
\PassOptionsToPackage{hyphens}{url}
\documentclass[
  12pt,
  a4paper,
]{article}
\usepackage{amsmath,amssymb}
\usepackage{lmodern}
\usepackage{setspace}
\usepackage{iftex}
\ifPDFTeX
  \usepackage[T1]{fontenc}
  \usepackage[utf8]{inputenc}
  \usepackage{textcomp} 
\else 
  \usepackage{unicode-math}
  \defaultfontfeatures{Scale=MatchLowercase}
  \defaultfontfeatures[\rmfamily]{Ligatures=TeX,Scale=1}
\fi
\IfFileExists{upquote.sty}{\usepackage{upquote}}{}
\IfFileExists{microtype.sty}{
  \usepackage[]{microtype}
  \UseMicrotypeSet[protrusion]{basicmath} 
}{}
\makeatletter
\@ifundefined{KOMAClassName}{
  \IfFileExists{parskip.sty}{%
    \usepackage{parskip}
  }{
    \setlength{\parindent}{0pt}
    \setlength{\parskip}{6pt plus 2pt minus 1pt}}
}{
  \KOMAoptions{parskip=half}}
\makeatother
\usepackage{xcolor}
\IfFileExists{xurl.sty}{\usepackage{xurl}}{} 
\IfFileExists{bookmark.sty}{\usepackage{bookmark}}{\usepackage{hyperref}}
\hypersetup{
  pdftitle={The Evolution of Scientific Credit: When Authorship Norms Impede Collaboration},
  pdfauthor={Toby Handfield and Kevin Zollman},
  hidelinks,
  pdfcreator={LaTeX via pandoc}}
\usepackage[top=20mm,left=25mm,right=25mm,heightrounded]{geometry}
\usepackage{graphicx}
\makeatletter
\def\maxwidth{\ifdim\Gin@nat@width>\linewidth\linewidth\else\Gin@nat@width\fi}
\def\maxheight{\ifdim\Gin@nat@height>\textheight\textheight\else\Gin@nat@height\fi}
\makeatother
\setkeys{Gin}{width=\maxwidth,height=\maxheight,keepaspectratio}
\makeatletter
\def\fps@figure{htbp}
\makeatother
\providecommand{\tightlist}{%
  \setlength{\itemsep}{0pt}\setlength{\parskip}{0pt}}
\usepackage{subcaption}
\ifLuaTeX
  \usepackage{selnolig}  
\fi

\usepackage[
  backend   = biber,
  style     = numeric,
  sorting   = none,
  sortcites = true,
  doi       = false,
  isbn      = false,
  url       = false,
  date      = year,
]{biblatex}

\DeclareFieldFormat[article]{journaltitle}{#1}
\DeclareFieldFormat[article]{title}{\mkbibquote{#1}}
\renewbibmacro*{in:}{}

\renewbibmacro*{journal+issuetitle}{%
  \usebibmacro{journal}%
  \setunit*{\addspace}%
  \usebibmacro{volume+number+eid}%
  \setunit{\addspace}%
  \printfield{year}%
  \setunit{\addcolon\space}%
  \usebibmacro{issue}%
  \newunit}

\AtEveryBibitem{\clearfield{note}\clearlist{language}}

\usepackage[sups]{XCharter}
\usepackage[charter]{mathdesign}
\DeclareMathAccent{\tilde}{\mathalpha}{operators}{"7E}

\title{The Evolution of Scientific Credit: When Authorship Norms Impede
Collaboration}
\author{Toby Handfield and Kevin
Zollman\thanks{Authors may or may not be listed in alphabetical order. TH: Monash University, KZ: Carnegie Mellon}}
\date{5 July 2025}

\begin{document}
\maketitle

\setstretch{1.3}
\frenchspacing

\begin{abstract}
  Scientific authorship norms vary dramatically across disciplines, from contribution-sensitive systems where first author is the greatest contributor and subsequent author order reflects relative input, to contribution-insensitive conventions like alphabetical ordering or senior-author-last. We develop evolutionary game-theoretic models to examine both how these divergent norms emerge and their subsequent effects on collaborative behavior. Our first model reveals that contribution-insensitive norms evolve when researchers who sacrifice positional advantage face the strongest adaptive pressure—for example senior authors managing larger collaboration portfolios or bearing heavier reputational stakes. This ``Red King'' dynamic potentially explains why fields in which senior researchers command large labs, major grants, and extensive collaboration portfolios may paradoxically evolve conventions that favour junior-author positioning. Our second model demonstrates that established norms influence researchers' willingness to collaborate, with contribution-sensitive norms consistently outperforming insensitive alternatives in fostering successful partnerships. Contribution-insensitive norms create systematic coordination failures through two mechanisms: ``main contributor resentment'' when exceptional work goes unrecognized, and ``second contributor resentment'' when comparable efforts receive unequal credit. These findings suggest that widely adopted practices like senior-last positioning and alphabetical ordering may function as institutional frictions that impede valuable scientific collaborations rather than neutral organizational conventions, potentially reducing overall scientific productivity across affected disciplines.
\end{abstract}

\hypertarget{the-evolution-and-effects-of-scientific-credit-attribution-norms}{%
\section{Introduction}\label{the-evolution-and-effects-of-scientific-credit-attribution-norms}}

Credit functions as the lifeblood of the scientific enterprise --
bylines, citation counts, and perceived ownership of ideas drive hiring,
promotion, grant awards, and career trajectories, thereby motivating
researchers to pursue ambitious questions and invest the effort
necessary for rigorous discovery \autocite{merton_sociology_1974}. At
the same time, collaboration has become the dominant mode of knowledge
production: multi-author papers now represent the lion's share of
publications, large consortia tackle complex problems, and empirical
studies link collaborative work to higher citation impact and greater
methodological innovation
\autocite{wuchty_increasing_2007,lariviere_team_2015,franceschet_collaboration_2011,bikard_exploring_2015,thelwall_why_2023}.
Yet reducing the rich tapestry of individual contributions -- ranging
from conceiving the research question and designing experiments to
collecting data, performing analyses, and drafting prose -- into a
simple, linear author list entails loss of information. That compression
can skew incentives and rewards, privileging certain roles or career
stages over others, and may shape both individual trajectories and the
collective pace of scientific progress in unexpected or undesirable ways.

To examine how different approaches to this information compression problem shape
scientific practice, we focus specifically on author-order norms, which
vary considerably across disciplines. Some fields employ
contribution-sensitive norms (C-Norms), where author order directly
signals relative contribution, typically with first authors having
contributed most. Other fields have adopted contribution-insensitive
norms (I-Norms), with perhaps the most notable being the alphabetical
norm often used in mathematics and economics. In many biological
sciences, there is a ``senior-author-last'' convention, whereby the last
author is understood to have had a supervisory role, and perhaps
provided funding or infrastructure, but with no determinate implication
regarding this author's contribution to the intellectual content of the
paper. The last author might have been the one who had the whole idea,
without which the paper wouldn't exist, or the last author might just be
coasting on the hard work and innovation of more junior colleagues.
Hence at least regarding the final position in the author list, this
norm is insensitive to contribution. Often the senior-author-last
convention is combined, however, with the presumption that first author
position signals principal contributor, while middle authors are listed
alphabetically, signaling roughly equal, but lesser contributions than
both first and last authors \autocite{mongeon_rise_2017}. Other
practices include footnoted equal-contribution statements, rotation
systems where collaborators take turns, and complex discipline-specific
conventions that blend multiple signals. For analytical clarity,
however, our investigation focuses on two-author collaborations,
allowing us to employ the simpler C-Norm versus I-Norm framework without
confronting these hybrid cases. While we primarily examine a senior-last
convention, our model generalizes to other contribution-insensitive
norms such as alphabetical ordering \autocite{waltman_empirical_2012}.

These relatively explicit, widely endorsed norms of author ordering of course interact with other, less official patterns of
credit assignment. The ``Matthew Effect,'' whereby
eminent scientists receive disproportionate credit for joint work
\autocite{merton_matthew_1968,perc_matthew_2014}, may lead readers to
attribute outsized credit to senior authors regardless of their position
in the author list. Similarly, there can be a bias towards giving more credit to first authors. Even in fields with alphabetical ordering,
readers might consciously know this convention carries no information
about contribution, yet still unconsciously attribute more credit to
earlier-listed authors
\autocite{weber_effects_2018,maciejovsky_research_2009}. Indeed,
surname-initial effects in disciplines with alphabetical norms create
strategic incentives and measurable efficiency losses in co-authorship
patterns
\autocite{ong_collaboration_2018,weber_effects_2018,efthyvoulou_alphabet_2008,einav_whats_2006,van_praag_benefits_2008}.
These informal biases operate alongside formal norms, creating tensions
between official conventions and unofficial credit attribution that can
undermine the informational value of authorship practices.

To disentangle these complex interactions between formal norms and
actual scientific practice, we need theoretical frameworks that can
account for both the emergence of different authorship conventions and
their subsequent effects on collaboration behaviors. To investigate these issues, we develop two
connected models. First, we analyze a simple game between two researchers who must agree on an authorship norm before
collaborating. This evolutionary model reveals when different norms are
likely to become established. Second, we examine how established norms
affect researchers' willingness to collaborate at all, identifying
conditions under which different norms promote or inhibit scientific
partnerships.

Building on the seminal analysis by Engers et al.
\autocite*{engers_firstauthor_1999}, which explored how authorship norms
affect effort allocation within collaborations, we shift attention to
prior stages: the formation of collaborations and the establishment of
norms themselves. Whereas Engers and colleagues analyzed how different
conventions influence individual effort once collaboration is underway,
we ask how collaborations arise in the first place and which norms are
likely to prevail. This shift reflects the sequential nature of
scientific activity: authors must first decide \emph{whether} to
collaborate and under \emph{which} convention, before deciding
\emph{how} to allocate effort. By foregrounding these upstream choices,
our models illuminate the broader structural conditions under which
individual strategic behavior occurs.

Our analysis yields two main results that illuminate the strategic foundations of scientific authorship. First, contribution-insensitive norms emerge not through egalitarian ideals but through asymmetric evolutionary pressures: these norms become entrenched when the researchers who sacrifice positional advantage (senior authors in senior-last systems, late-alphabet researchers in alphabetical ordering) face the most intense adaptive pressure. Whether driven by managing larger collaboration portfolios, bearing heavier reputational consequences, or confronting steeper opportunity costs from partnership failures, these high-stakes actors evolve toward accommodating conventions that formally disadvantage them. This finding reveals that ostensibly neutral practices like alphabetical ordering or senior-last positioning actually encode the strategic adaptations of those with the most to lose, not democratic consensus or historical accident.

Second, this evolutionary logic produces systematic inefficiency: contribution-sensitive norms consistently outperform their insensitive counterparts in fostering successful collaborations. The compression of rich contribution information into positionally arbitrary signals creates predictable coordination failures, suggesting that disciplines employing senior-last or alphabetical conventions may be systematically discouraging valuable scientific partnerships. These widely adopted practices appear to function as institutional frictions that impede collaborative science rather than neutral organizational conventions.

These findings connect to broader investigations of how scientific norms
shape knowledge production
\autocite{Kitcher1993-KITTAO-2,fortunato_science_2018,ioannidis_meta-research_2015}.
Science operates through an intricate web of formal and informal
conventions: statistical thresholds that determine significance,
methodological standards for gathering evidence, what constitute
interesting or important problems \autocite{foster_tradition_2015},
expectations around peer review, criteria for promotion, and of course,
protocols governing authorship
\autocite{barabasi_credit_2021,jian_perceptions_2013,mongeon_rise_2017,marusic_systematic_2011}.
Although these norms often present as mere traditions or accidents of
history, they reflect complex interactions between incentive structures,
power dynamics, and information needs. The norms and customs of
scientific practice -- including authorship conventions -- continuously
evolve through competitive adaptation, strategic behavior, and
institutional path-dependencies.

Viewed through this lens, our results offer important insights for both
understanding and potentially reforming scientific institutions. Credit
attribution systems serve dual roles: they function as incentive
mechanisms that motivate effort and information channels that signal
contribution. Our findings reveal how tensions between these roles can
reduce scientific productivity and potentially skew career trajectories.
By treating authorship norms as strategic choices with field-wide
consequences, our approach provides a framework for analyzing the
evolution and effects of scientific conventions more broadly, suggesting
that seemingly neutral organizational practices may significantly
influence both individual careers and collective knowledge generation.

\hypertarget{model-and-results}{%
\section{Model and Results}\label{model-and-results}}

To explore how author order norms arise and influence scientific
collaboration, we develop two connected models. These models analyze
simplified games that researchers ``play'' when deciding how to
collaborate and how to assign authorship credit. We focus on two main
types of norms: \textbf{contribution-sensitive norms (C-Norms)}, where
author order reflects who contributed more, and
\textbf{contribution-insensitive norms (I-Norms)}, like alphabetical
ordering or senior-author-last, where order doesn't signal contribution.

\hypertarget{how-author-order-norms-emerge-a-simple-game}{%
\subsection{How Author Order Norms Emerge: A Simple
Game}\label{how-author-order-norms-emerge-a-simple-game}}

Our first model examines how contribution-sensitive norms (C-Norms) and
contribution-insensitive norms (I-Norms) might become established in
scientific fields. We consider a simplified scenario with two
researchers -- ``Junior'' and ``Senior'' -- who are contemplating
collaboration. Before they begin working together, they must implicitly
agree on which norm to use if they publish together.

Each researcher proposes a norm: either the C-Norm or the I-Norm. The
C-Norm places the researcher who contributed more as first author, while
the I-Norm always places Junior first. For collaboration to proceed,
these proposals must be compatible. ``Compatible'' means they don't
anticipate any irresolvable disagreement over who gets the \emph{first}
author position when publication time arrives. Compatibility arises in
three scenarios: (1) both choose the C-Norm, (2) both choose the I-Norm,
or (3) Senior chooses the I-Norm while Junior chooses the C-Norm. This
last case represents a situation where both researchers are effectively
yielding priority to the other. Senior is saying ``you should be listed
first regardless''; while Junior is saying ``no, if you contribute more,
you should go first''. We assume they resolve this amicable standoff by
flipping a coin to determine which norm to follow. The only incompatible
scenario, preventing collaboration entirely, occurs when Junior insists
on the I-Norm (guaranteeing their first authorship) while Senior insists
on the C-Norm (where they would be first if they contribute more).

If the researchers successfully agree on a norm and collaborate, the
proportion contributed by each is determined by an exogenous
distribution. We characterize this distribution using three parameters:
the probability that Junior contributed more (\(w_j\)), Junior's
expected contribution given that they contributed more (\(b_j\)), and
Senior's expected contribution given that they contributed more
(\(b_s\)).

When deciding which norm to adopt, researchers consider what maximizes
their expected \emph{credit}. This credit is assigned by a third party,
representing the scientific community, who observes the published paper
and attempts to determine how much each author contributed, based on:
the observed author order, their prior expectations about the relative
contributions of the authors, and their understanding of prevailing
norms (i.e.~what the author's mixed strategies are in the game).

We assume authors bargain over author order before they know the
magnitude of their individual contributions. At time of decision, they
merely know the probability distribution over relative contributions
(\(w_j, b_j, b_s\)), and they know the value that would be added to
their ideas if they are able to agree on an author order and publish
those ideas together (\(\hat{c}\)).

Formal norms alone don't determine credit assignment; informal biases inevitably intrude. We model this with two parameters. First, ($\varepsilon > 0$) represents the probability that the community mistakenly gives all credit to the first author, even under an I-Norm where position ostensibly carries no signal about contribution. This models the stubborn tendency to remember and cite first authors regardless of field conventions. Second, we introduce ($\chi > 0$) to capture the classic ``Matthew Effect'' \autocite{merton_matthew_1968}, whereby with probability ($\chi$) the senior author receives all credit regardless of author order or actual contribution. These parameters reflect the cocktail of cognitive biases, information shortcuts, and social dynamics that shape scientific recognition beyond formal norms\autocite{sarsons_recognition_2017,weber_effects_2018,barabasi_scientific_2021,van_praag_benefits_2008,einav_whats_2006}.

So if the authors agree on compatible norms, they collaborate, and they
together realize a paper of value \(1 + \hat{c}\), where \(\hat{c} > 0\)
is a parameter reflecting the additional value generated by
collaboration. The scientific community observes the value of the joint
paper perfectly, and then makes an estimate, conditional on their beliefs
about the players' mixed strategies (they know what norm is likely to be
in effect) and on the scientists' expected contributions (they have
accurate beliefs about the relative abilities of the individuals), of
the individual contributions of each scientist. These credit
attributions by the community simply are the payoffs to the authors.

If the authors do not agree on norms then they must publish their ideas
separately. The community is then able to precisely estimate who did
what, and the payoff to each is simply the value of their respective
contributions, without the benefit of collaboration.

Effectively, this is a 2 × 2 coordination game, but with a distinctive
twist: the exact payoffs depend on the strategies the authors are
perceived to be using. The scientific community's assessment of each
contributor's work -- and thus the credit allocated -- hinges on their
beliefs about which norms are in operation among scientists. This
creates a feedback loop between strategy choices and payoff structures
that standard game matrices cannot easily capture.

Despite this complexity, we can identify clear preference rankings.
Junior authors generally prefer the outcome where both adopt the I-norm
$(I,I)$ over the scenario where both use the C-norm $(C,C)$. Senior authors,
conversely, prefer $(C,C)$ over $(I,I)$. This preference divergence occurs
not because of differences in the community's accurate credit assessment
-- in both equilibria, when the community correctly identifies the
operative norm, authors receive credit matching their expected
contributions. Rather, the conflict stems from our parameter
\(\varepsilon\), which captures first-author bias: under the I-norm,
Junior always enjoys this positional advantage, while under the C-norm,
Senior maximizes their opportunities to be listed first and capture this
bonus.

Both authors have a common interest in avoiding the incompatible
combination $(C,I)$, where Senior insists on contribution-based ordering
while Junior demands the I-norm. This scenario alone prevents
collaboration entirely, making it mutually disadvantageous compared to
either coordinated outcome.

Using replicator dynamics to analyse the model, both C-Norms and I-Norms
can emerge as stable equilibria, with no other stable equilibria
existing under realistic conditions of first-author bias (see
Supplementary Information for discussion of special cases where first
author bias is extraordinarily large). However, the likelihood of each
norm becoming established depends on the distribution of contributions.
In Figure 1 we show a stream plot of the replicator dynamics, for three
different initial distributions of expected contributions by the
authors.

\begin{figure}
\centering
\includegraphics{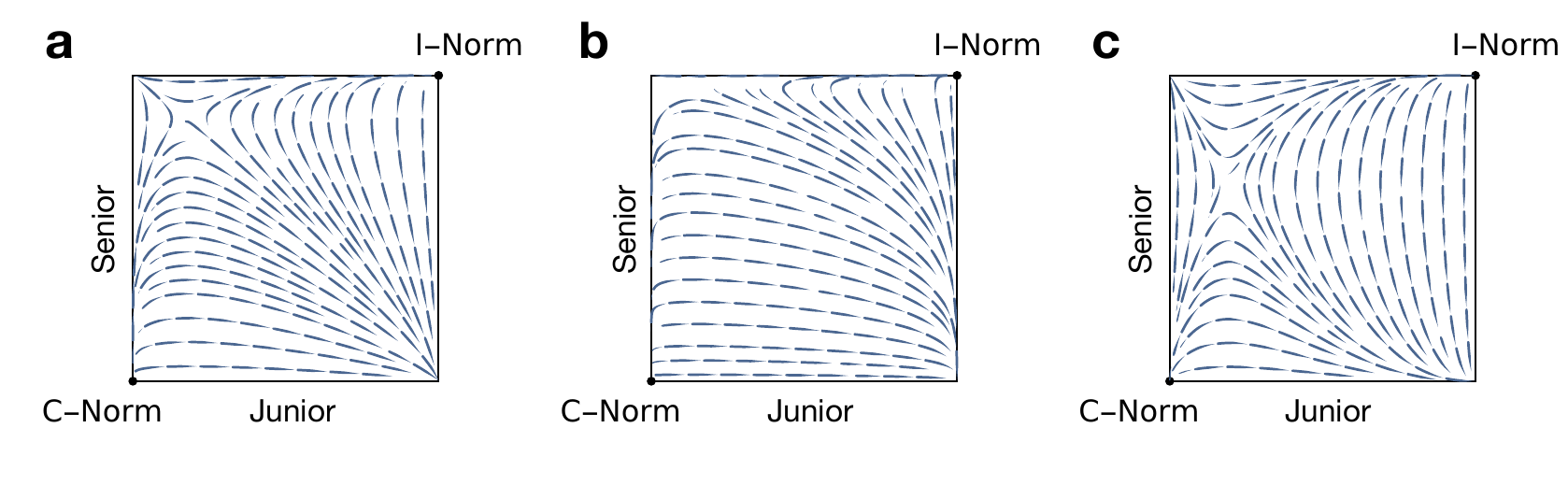}
\caption{\textbf{Evolution of authorship norms under different prior
expectations of contribution.} Phase-plane diagrams
showing replicator dynamics for the two-population game. Each streamline
indicates the trajectory of norm adoption, with comet heads pointing
toward evolutionary outcomes. The horizontal axis represents the
proportion of junior researchers adopting the I-norm. The vertical axis shows the
proportion of senior researchers adopting the I-Norm. Solid dots mark stable
equilibria, located at opposite corners: bottom-left (all researchers
follow C-Norm) and top-right (all researchers follow I-Norm). Community
probability of incorrectly attributing credit to first author regardless
of norm: \(\varepsilon = 0.1\). Community probability of attributing
credit to senior author regardless of norm: \(\chi = 0.05\).
Each stream plots is derived from a different underlying probability
distributions for prior expectations about relative
contributions of the researchers. (\textbf{a}):
symmetric distribution (equal chances for either to contribute more);
(\textbf{b}): junior-biased distribution; (\textbf{c}): senior-biased distribution. 
When seniors are expected to contribute more (c),
the I-Norm's basin of attraction expands, making the senior-last
convention more likely to emerge. Conversely, when juniors are
expected to contribute more (b), the C-Norm's basin widens, favoring
the greatest-contributor-first convention. With symmetric
contribution expectations (a), neither norm has a decisive evolutionary
advantage. This pattern remains consistent across various
distribution shapes, and for a range of values of \(\varepsilon, \chi\).
See SI for illustrations.}
\end{figure}

The most striking finding from this first model is that the I-Norm
(senior-author-last) is more likely to evolve when the Senior author is
expected to make a greater contribution than Junior. Conversely, when
Junior author is expected to make the greater contribution, the C-norm
is more likely to evolve. This occurs because when one author is
expected to make a greater contribution, they stand to lose more from a
failed collaboration: this entails that they are subject to a greater
selection pressure, and thus adapt their strategies more rapidly than
the other author. For instance, when Senior faces higher stakes in
successful collaboration, they evolve relatively fast toward strategies
that maximize the probability of collaboration occurring at all -- in
this case, choosing the I-Norm, which is compatible with either of
Junior's possible strategies. Conversely, when Junior faces higher
stakes in successful collaboration, they evolve relatively fast towards
the C-norm, which is compatible with either of Senior's possible
strategies.

This dynamic extends beyond asymmetries in expected contributions. The
key determinant is which party faces greater evolutionary pressure in
aggregate. While junior academics are typically in less secure
employment, and this is a very salient reason to think of them as under
greater adaptive pressure, this is ultimately an empirical question that
may well vary across fields. Several considerations might counterbalance
the employment security factor. (i) Senior researchers typically manage
a larger portfolio of simultaneous collaborations, exposing them to
selection pressures across multiple fronts. (ii) Each publication often
carries heavier consequences for senior researchers, whose reputations,
grant renewals, and lab sustainability depend on consistent output.
(iii) Junior researchers paradoxically benefit from more viable outside
options and higher mobility between career paths, potentially reducing
their sensitivity to academic selection. (iv) Senior authors might be
expected to make the greatest contribution to any collaboration, whether
from accumulated expertise, a reputation for brilliance, or the
gravitational pull of the Matthew Effect, and this makes the opportunity
cost of a failed collaboration particularly costly for senior
researchers. The relative strength of these countervailing forces
presumably varies across disciplines and institutions, and their
empirical weight remains an open question.

This evolutionary mechanism, through these channels or others, might
explain some of the diversity of authorship norms across scientific
disciplines. Fields where senior authors typically contribute more or
collaborate at a higher rate should, according to our model, more
frequently develop senior-author-last conventions. This prediction
aligns with observed patterns in laboratory sciences where principal
investigators oversee multiple research teams simultaneously, and where
senior-author-last appears to be a more common norm. Conversely, fields
where junior and senior researchers maintain more equal collaboration
loads and contribution expectations should more frequently develop
contribution-sensitive norms.

\begin{figure}[t]
\centering
\includegraphics{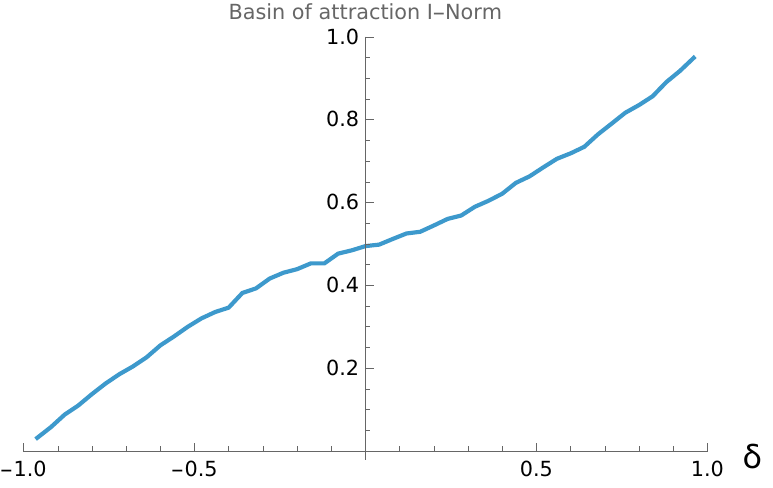}
\caption{\textbf{Basin of attraction for the I norm, as a function of difference in expected contributions between senior and junior author.} \(\varepsilon = 0.1\, \chi = 0.05\). The qualitative result is
robust for various levels of \(\hat{c}, \varepsilon, \chi\), see SI.}
\end{figure}

This finding is an example of what evolutionary game theorists call the
``Red King effect'' \autocite{bergstrom_red_2003}. Contrary to the more
familiar Red Queen dynamics where faster evolution is advantageous, the
Red King effect describes situations where slower-evolving populations
can achieve more favorable outcomes in mutualistic interactions. When
Senior is expected to contribute more, they have more to lose from
failed collaborations, creating stronger selective pressure on their
strategies. This accelerated evolution paradoxically leads Senior to
adopt the more accommodating I-Norm strategy that guarantees
collaboration but sacrifices potentially deserved first-author credit.
Junior, evolving more slowly due to lower stakes, maintains the C-Norm
preference longer, eventually pulling the system toward that
equilibrium. This aligns with Bruner's \autocite{bruner_minority_2019}
findings on bargaining norm evolution, where faster-adapting groups
often settle for less favorable divisions. Our model thus reveals how
asymmetric evolutionary speeds -- driven by differential stakes in
scientific collaboration -- can shape authorship conventions across
disciplines, potentially explaining why fields with powerful senior
researchers might counterintuitively develop norms that favor junior
authors in author ordering.

All of our conclusions about this first model carry over unchanged if we reinterpret the I-norm as simply alphabetical ordering, treating our two researchers Junior and Senior as ``A" and ``B", respectively. In this guise, the replicator-dynamic equations and basins of attraction are identical: alphabetical ordering (the I-norm) becomes more likely to evolve whenever authors with later-alphabet names face greater selective pressure than those earlier in the alphabet. However, unlike seniority, which correlates with systematic differences in collaboration patterns, career stakes, and research portfolios, alphabetical position is essentially random and unlikely to create the persistent asymmetries in adaptive pressure that our model requires. Authors named ``Smith'' or ``Zhang'' don't systematically face higher collaboration costs or reputational risks than those named ``Anderson'' or ``Brown.'' This suggests that while our framework demonstrates the theoretical possibility of alphabetical norms emerging through evolutionary dynamics, their actual adoption likely stems from different mechanisms -- perhaps coordination convenience or historical accident -- rather than strategic adaptation. What does persist, however, is the key takeaway of the first model: both contribution-sensitive and contribution-insensitive norms are stable evolutionary equilibria, and so the C-norm is far from an inevitable convention.

\hypertarget{how-norms-affect-collaboration-will-they-even-work-together}{%
\subsection{How Norms Affect Collaboration: Will They Even Work
Together?}\label{how-norms-affect-collaboration-will-they-even-work-together}}

Our second model shifts focus to the consequences of having either a
C-Norm or an I-Norm already in place. We assume a norm is established in
a field and want to know how these different norms affect researchers'
willingness to collaborate.

Again considering Junior and Senior, they now know what the authorship
norm in their field is. After collaboration, they will know exactly how
much each contributed. The question is: will they choose to publish
jointly, or would one of them rather ``go it alone'' and publish their
part separately?

If there were no benefit to collaboration, then the decision to collaborate would be strictly zero sum: if one author gets assigned more credit by the community than their actual contribution, that author benefits from collaboration, while the other will lose, making collaboration inevitably fail. However, because we assume that collaboration improves the quality of the work -- either by boosting visibility or by generating better ideas \autocite{thelwall_why_2023,bikard_exploring_2015} -- the game is non-zero sum and there is the potential for both authors to prefer collaboration, even though the community will almost inevitably give one author less credit than their actual share.

Credit is assigned by the same process as in model 1: the community has
priors on the contributions of both authors, and also knows the norm in
place. The community observes the quality of the paper and, conditional
on its priors, assigns credit to each author.

Authors anticipate being judged for their contribution by the community,
and can thus estimate whether their payoff will be greater if they ``go
it alone'' or if they collaborate. The collaboration will only proceed
if both authors prefer that option. The important question then, is when
will authors refuse to collaborate under each norm.

\textbf{Collaboration Under the I-Norm:} With the I-Norm, collaborations
will break down when one author contributes substantially more than was
expected (``main contributor resentment''). This is because the I-norm
will cap the credit a scientist receives at whatever the community
antecedently expected. If either scientist makes a surprisingly large
contribution, the I-norm cannot reflect this, and this can lead the
author to prefer sole-authorship. See Figure 2, left column.

\textbf{Collaboration Under the C-Norm:} The C-Norm presents different
challenges. C-Norms can discourage collaborations in two situations
(Figure 2, right column):

\begin{enumerate}
\def\labelenumi{\arabic{enumi}.}
\tightlist
\item
  \textbf{When contributions are very unequal:} Being first author sends
  only a very broad signal, that may not give sufficient reward when one
  author has made an outsize contribution. If one person does almost all
  the work, they will receive more credit by publishing alone than
  they'll receive from being first author under a C-Norm. This is
  similar to the way collaborations fail under the I-norm (``main
  contributor resentment'' again), but because the C-norm sends a
  stronger signal of who contributed the most, it generally enables
  collaboration under a wider range of unequal contributions.
\item
  \textbf{When contributions are very similar:} Under a C-Norm, if
  contributions are almost equal, whoever ends up listed second might be
  under-credited compared to what they could achieve by publishing
  independently (``second contributor resentment''). This occurs because
  the C-Norm forces a fixed distinction in credit, even when
  contributions are nearly equivalent.
\end{enumerate}

In general, collaborations fail less often when the value of
collaboration (\(\hat{c}\)) is high. This is straightforward to explain:
when the benefits of cooperating are greater, it is easier to achieve
cooperation, because it is more likely to be in the interests of both.

\begin{figure}
\centering
\includegraphics{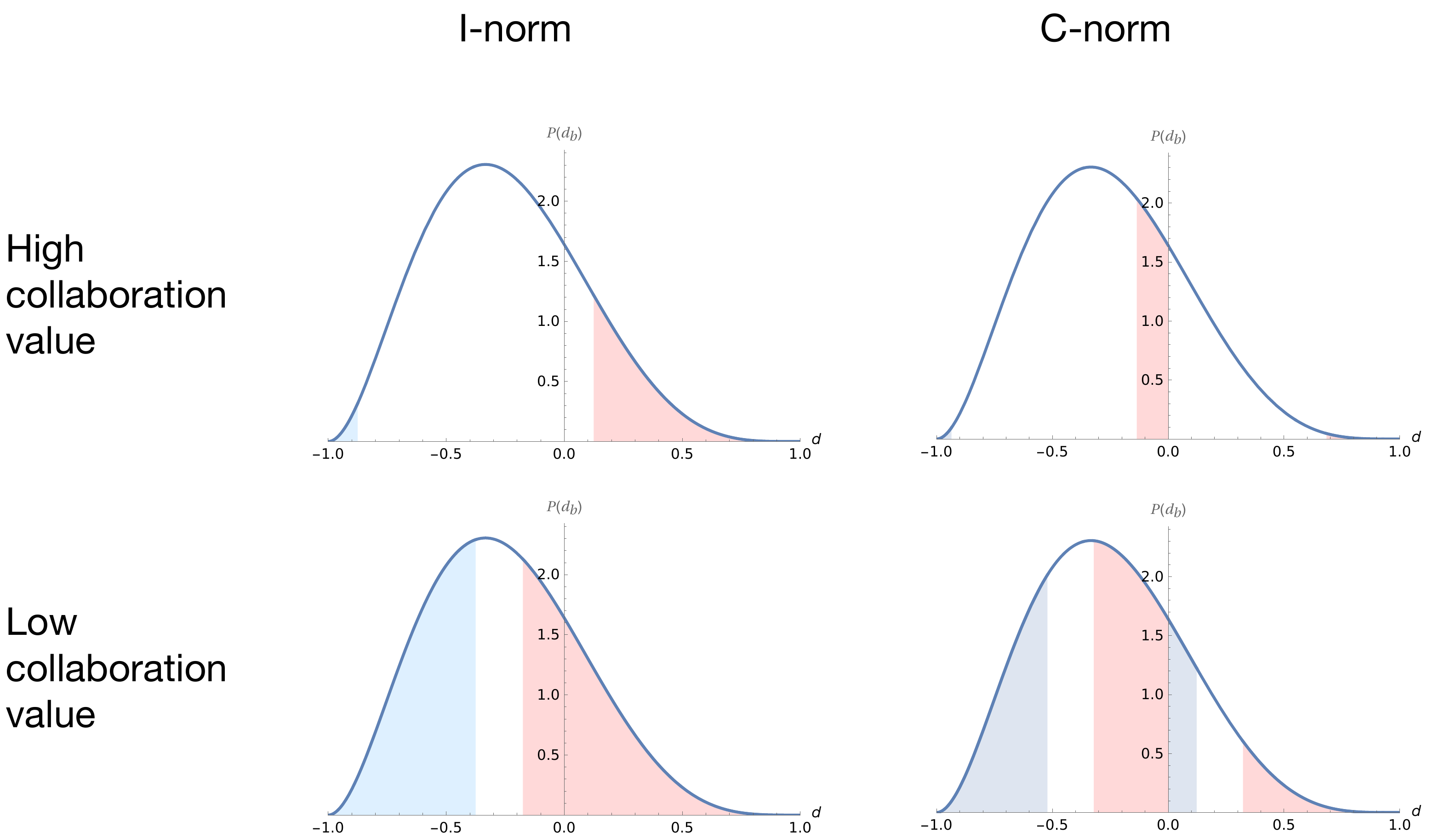}
\caption{\textbf{Collaboration failure regions under different authorship norms.}
The horizontal axis represents the ex-post difference in contributions
between Junior and Senior (\(d = c_S - c_J\)), while colored regions
indicate where collaboration breaks down (red: senior refuses; blue:
junior refuses). The bell-shaped curves illustrate probability
distributions of contributions. Under the I-Norm (left panels),
collaborations primarily fail at the extremes due to ``main contributor
resentment'' -- when one author contributes substantially more than
expected but receives insufficient credit. Under the C-Norm (right
panels), failures occur both at extremes and in the middle zone of
near-equal contributions, where ``second contributor resentment''
emerges because the norm forces a stark credit distinction despite
similar work. Upper panels show results with high collaboration value
(\(\hat{c} = 0.05\)), while lower panels show low collaboration value
(\(\hat{c} = 0.01\)).}
\end{figure}

\textbf{Comparing the Norms:} Neither norm is perfect. Both can lead to
collaboration failures. However, when comparing them across various
contribution scenarios, we find that the I-Norm tends to lead to more
collaboration failures and a greater loss of scientific value from
collaboration. The C-Norm, while imperfect, generally fostered more
successful collaborations in our model. See Figure 4. There are some
cases where the I-norm is superior -- and this is generally where the
collaborations are of moderate to high value, and the expected
contributions are highly unequal.

\begin{figure}
\centering
\includegraphics[width=0.9\textwidth,height=\textheight]{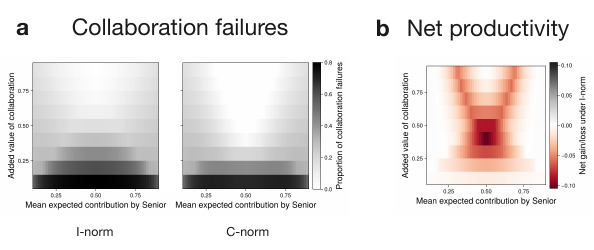}
\caption{\textbf{Comparative efficiency of
contribution-sensitive versus contribution-insensitive norms.}
\textbf{Panel A:} Collaboration failure rates for the C-norm and the
I-norm across parameter space. Darker regions indicate higher failure
rates. \textbf{Panel B:} Net efficiency comparison between norms
(difference between the norms in collaboration failures, weighted by
collaboration value). Negative values (red) indicate the C-norm is more
efficient; positive values (grey) indicate the I-norm is more efficient.
The C-norm demonstrates superior efficiency across most parameter
combinations, with the I-norm showing only marginal advantages in
limited scenarios (faint grey regions). The underlying probability
distributions are all beta distributions with \(\alpha + \beta = 7\).
The horizontal axis represents Junior's expected contribution
(\(\mu_j\)), while the vertical axis shows collaboration value
(\(\hat{c}\)). Additional distribution scenarios illustrated in
Supplementary Information.}
\end{figure}

As with the first model, while our primary analysis has focused on the
senior-author-last convention, these results generalize meaningfully to
other contribution-insensitive norms, most notably alphabetical
ordering. Just as with the senior-last norm, alphabetical ordering fails
to precisely communicate actual contributions, creating similar
inefficiencies in collaboration decisions. When author order is
determined alphabetically, researchers with greater contributions --
regardless of their surname's position in the alphabet -- may find their
work undervalued and choose to publish independently rather than
collaborate. This suggests that disciplines with alphabetical ordering
conventions like economics, mathematics, and theoretical computer
science may experience collaboration failures similar to those in fields
using senior-last conventions. Recent empirical work has documented
strategic behaviors and publication patterns consistent with these
predicted inefficiencies, including a greater reluctance to collaborate
on the part of authors with surnames late in the
alphabet \autocite{weber_effects_2018,ong_collaboration_2018}. The
potential lost scientific value in fields using alphabetical norms could
be significant, particularly for collaborative projects where the
benefits of teamwork are modest. Just as our model predicts for
senior-last norms, contribution-sensitive alternatives might foster more
effective collaboration across disciplines currently relying on
alphabetical conventions.

\hypertarget{discussion}{%
\section{Discussion}\label{discussion}}

Our analysis demonstrates two key findings about scientific authorship
norms. First, both contribution-sensitive norms and
contribution-insensitive norms can emerge as stable equilibria in
scientific communities, with the likelihood of each norm becoming
established dependent on field-specific factors. Critically, we found
that contribution-insensitive norms like the senior-author-last
convention are more likely to evolve when senior researchers face
greater evolutionary pressure, such as when they engage in multiple
simultaneous collaborations or when they have higher stakes for failed
collaborations.

Second, and perhaps more importantly, we found that different authorship
norms have significant implications for scientific collaboration. While
both C-Norms and I-Norms can lead to collaboration failures under
certain conditions, contribution-sensitive norms generally foster more
successful collaborations across a wider range of scenarios. When
researchers know their actual contributions will be reflected in author
ordering, they are more likely to pursue valuable joint work.

These findings have important implications for understanding scientific
productivity and potentially for reforming scientific institutions. The
norm dynamics we identify may help explain the puzzling diversity of
authorship conventions across scientific disciplines. Moreover, our
results suggest that fields employing contribution-insensitive norms
might be inadvertently discouraging collaborations that would produce
valuable scientific knowledge.

\hypertarget{comparison-with-existing-literature}{%
\subsection{Comparison with Existing
Literature}\label{comparison-with-existing-literature}}

Our evolutionary framework for analyzing authorship norms provides
several distinct contributions over previous approaches. Engers et al.
\autocite{engers_firstauthor_1999} pioneered formal modeling of
authorship conventions, examining a scenario where researchers negotiate
author order after observing their contributions. Their model employs
Nash bargaining to resolve authorship disputes and yields a striking
conclusion: contribution-insensitive norms (alphabetical ordering in
their preferred interpretation) constitute stable equilibria, while
contribution-sensitive norms never achieve equilibrium status. This
prediction, however, creates an immediate puzzle given the widespread
prevalence of contribution-sensitive norms across numerous scientific
disciplines.

This predictive limitation stems partly from their methodological
choices. The axiomatic bargaining approach, while elegant, assumes
complete information, simultaneous choice, and perfectly rational actors
operating with common knowledge of the bargaining structure. Scientific
norm evolution, by contrast, emerges through iterative interactions
among heterogeneous agents (from graduate students to senior
researchers) with varying information access and bargaining power,
operating in environments where conventions develop gradually through
path-dependent processes rather than clean-slate negotiations. Our
evolutionary dynamics framework better captures these messy realities of
scientific practice, explaining how both norm types can emerge as stable
equilibria depending on field-specific conditions and selection
pressures.

The models also differ in their analytical focus. Engers and colleagues
endogenize effort decisions while treating collaboration as inevitable,
investigating how different norms influence researchers' incentives to
contribute optimally to joint work. We take the complementary approach
of making contribution levels exogenous while endogenizing the
collaboration decision itself. This shift in emphasis allows us to
explore when valuable scientific partnerships form at all, rather than
assuming they always materialize. Both perspectives offer valuable
insights into different aspects of scientific practice -- one addressing
effort allocation within established collaborations, the other examining
the upstream decision to collaborate in the first place.

Notably, these distinct approaches converge toward similar conclusions
about norm efficiency. Engers et al.~find that contribution-insensitive
norms provide inferior incentives for optimal effort allocation compared
to contribution-sensitive alternatives. Our analysis reveals that
contribution-insensitive norms also create more frequent collaboration
failures than their contribution-sensitive counterparts. This
convergence from different analytical angles strengthens the case that
contribution-sensitive norms may better serve both individual
researchers and scientific productivity as a whole.

Bikard et al. \autocite{bikard_exploring_2015} offer another important
comparison point, investigating how credit allocation shapes
collaboration choices under fixed attribution rules. Their empirical
finding that collaboration yields individual benefits primarily when
credit is disproportionately allocated reinforces our theoretical
insight that authorship norms function simultaneously as incentive
mechanisms and information channels. Our work extends their analysis by
endogenizing the conventions themselves, demonstrating how norms emerge
as equilibrium outcomes of strategic interactions among researchers with
divergent interests. This evolutionary perspective adds theoretical
foundation to empirical observations about scientific collaboration
patterns, while suggesting new avenues for empirical research on how
credit-allocation practices shape scientists' collaboration decisions
across different fields.

\hypertarget{limitations-and-future-directions}{%
\subsection{Limitations and Future
Directions}\label{limitations-and-future-directions}}

Our models have several limitations that might be addressed in future
work.

First, our model of collaboration failures focuses specifically on
credit attribution issues, but there are other reasons why scientific
collaborations might fail. Coordination challenges, differing research
priorities, and resource constraints can all impede joint work. However,
these factors can be conceptually incorporated into our parameter for
the benefit of collaboration (\(\hat{c}\)). When these additional
obstacles are significant, the effective value of collaboration
decreases, potentially leading to the patterns of collaboration failure
we identify.

Second, our analysis considers only two researchers and two possible
norms. Extending the model to include larger collaborations and a
broader range of potential norms would provide additional insights. This
extension would be particularly valuable given empirical findings that
authorship patterns become more complex with more authors
\autocite{correa_jr_patterns_2017}, and that different disciplines
exhibit distinct patterns of labor division across authorship positions
\autocite{lariviere_contributorship_2016}. Including contribution
statements or equal contribution designations in our models could
capture more nuanced aspects of modern scientific collaboration.

Third, while our model demonstrates the theoretical efficiency
advantages of contribution-sensitive norms, it does not address the
practical challenges of implementing such norms in fields that have
established conventions. Norm transitions involve coordination problems
and potential resistance from researchers who benefit from existing
arrangements.

\hypertarget{implications-for-scientific-practice}{%
\subsection{Implications for Scientific
Practice}\label{implications-for-scientific-practice}}

Our analysis suggests potential benefits from contribution-sensitive
norms, though translating these theoretical insights into practical
recommendations requires careful consideration. While our models provide
a coherent framework for understanding authorship conventions, they
represent just one -- very preliminary -- approach to a complex social
phenomenon. Any proposed reforms to established practices should be
supported by corroborating evidence -- both empirical studies across
disciplines and complementary theoretical models -- before we can begin
to estimate the likely effects of interventions. The limitations of our
approach highlight the importance of triangulating findings through
multiple methodologies before advocating for systematic changes to
scientific credit attribution systems.

In real scientific communities, senior figures derive authority and
reputational capital from existing byline rituals and are unlikely to
cede status absent compelling incentives. Moreover, mandating detailed
contribution statements risks devolving into perfunctory bureaucracy --
boxes ticked for compliance, not clarity -- unless accompanied by robust
enforcement and accountability structures. Any credible effort to reform
authorship conventions must therefore rest on realistic assumptions
about strategic behavior, clear protocols that deter token compliance,
and a foundation of evidence extending well beyond a single model.

\hypertarget{conclusion}{%
\subsection{Conclusion}\label{conclusion}}

Scientific authorship norms are not merely conventional practices but
structural features of scientific communities that shape collaboration
patterns and knowledge production. Our evolutionary approach to
understanding these norms reveals how seemingly arbitrary conventions
can arise from asymmetric selection pressures and how they subsequently
affect scientific collaboration.

The efficiency advantages of contribution-sensitive norms identified in
our analysis suggest that scientific communities should critically
examine their authorship practices. By aligning credit attribution more
closely with actual contributions, fields might foster more valuable
collaborations and ultimately enhance scientific progress. While
changing established norms presents significant challenges, the
potential benefits for scientific productivity make such efforts worthy
of consideration.

Future work should extend our models to more complex collaboration
scenarios, investigate empirical patterns of authorship and
collaboration across different fields, and explore practical
interventions that might shift credit attribution practices toward more
efficient arrangements. By treating authorship norms as evolving social
conventions rather than fixed traditions, we open new possibilities for
understanding and potentially improving the social structure of science.

\hypertarget{methods}{%
\section{Methods}\label{methods}}

\hypertarget{model-1-the-evolution-of-authorship-norms}{%
\subsection{Model 1: The Evolution of Authorship
Norms}\label{model-1-the-evolution-of-authorship-norms}}

To analyze how different authorship norms emerge, we developed a
game-theoretic model with two researchers whom we call Junior and
Senior. Before collaboration, each researcher proposes either a
contribution-sensitive norm (C-Norm), where author order reflects
relative contribution, or a contribution-insensitive norm (I-Norm),
where Junior is always listed first regardless of contribution.

Collaboration proceeds if their norm proposals are compatible.
Compatibility occurs when: (1) both choose the C-Norm, (2) both choose
the I-Norm, or (3) Senior chooses the I-Norm while Junior chooses the
C-Norm. In the last case, we assume they determine which norm to use by
flipping a coin. The only incompatible scenario occurs when Junior
insists on the I-Norm while Senior demands the C-Norm, leading to no
collaboration.

If collaboration proceeds, the value of their joint paper is
\(1 + \hat{c}\), where \(\hat{c} > 0\) represents the added value from
collaboration. Nature determines each researcher's contribution, with
Junior's contribution denoted as \(c_j\) and Senior's as \(1-c_j\). The
probability that Junior is the greater contributor is \(w_j \in (0,1)\),
with expected contribution \(b_j \in (0.5,1)\) if this is the case.
Similarly, Senior's expected contribution when they contribute more is
\(b_s \in (0.5,1)\).

If researchers fail to agree on a norm, they publish independently, with
papers of value \(c_j\) and \(1-c_j\) respectively. In this case, the
scientific community (modeled as an unbiased third party) assigns credit
perfectly aligned with actual contributions.

When researchers collaborate, credit assignment becomes more complex.
The scientific community observes author order and, knowing the
researchers' strategies and the underlying probability distribution,
estimates each researcher's contribution. We model a small probability
\(\varepsilon > 0\) that the community mistakenly gives all of the
credit to the first author, regardless of the community's norms. We also
include a small probability \(\chi > 0\) that the community mistakenly
gives all the credit to the senior author.

For example, under the I-Norm (Junior always first), if the community
observes Junior listed first, they calculate the probability that Junior
actually contributed more as:

\[m_j = \frac{(1 - p_j(1 - p_s))w_j}{(1 - p_j(1 - p_s))w_j + (p_j p_s + 0.5(1 - p_j)p_s)(1 - w_j)}\]

where \(p_j\) and \(p_s\) are the probabilities that Junior and Senior
play the I-Norm, respectively. Junior's expected credit is then
\(m_j b_j + (1-m_j)(1-b_s)\).

We analyze this game using replicator dynamics to model how strategies
evolve in response to payoffs. The replicator dynamics for our
two-population game are:

\[\dot{p}_j = p_j(1-p_j)(\pi_j(I\text{-Norm}) - \pi_j(C\text{-Norm}))\]
\[\dot{p}_s = p_s(1-p_s)(\pi_s(I\text{-Norm}) - \pi_s(C\text{-Norm}))\]

where \(\pi_j\) and \(\pi_s\) are the expected payoffs for Junior and
Senior when playing each strategy.

\hypertarget{model-2-collaboration-decisions-under-established-norms}{%
\subsection{Model 2: Collaboration Decisions Under Established
Norms}\label{model-2-collaboration-decisions-under-established-norms}}

Our second model examines how established norms affect researchers'
willingness to collaborate. We assume a norm is already in place and
researchers know their exact contributions before deciding whether to
publish jointly or independently.

If they publish independently, each receives credit equal to their
contribution (\(c_j\) and \(1-c_j\)). If they collaborate, their joint
paper has value \(1 + \hat{c}\). The scientific community assigns credit
based on author order and the established norm. Researchers collaborate
only if both weakly prefer this option to publishing alone.

Under the I-Norm, collaboration fails when one researcher contributes
substantially more than expected. This occurs when:

\begin{itemize}
\item
  Junior refuses if \(c_j > \mu_j(1+\hat{c})\), where \(\mu_j\) is the
  community's prior expectation of Junior's contribution
\item
  Senior refuses if \(1-c_j > (1-\mu_j)(1+\hat{c})\)
\end{itemize}

Under the C-Norm, collaboration can fail in two distinct regions:

\begin{enumerate}
\def\labelenumi{\arabic{enumi}.}
\item
  When contributions are highly unequal: the main contributor refuses if
  they did substantially more work than would be signaled by first
  authorship;
\item
  When contributions are nearly equal: the second-listed author refuses
  if their contribution is close to the first author's but they would
  receive disproportionately less credit.
\end{enumerate}

For each norm, we calculate the ex-ante probability of collaboration
failure and the expected loss of scientific value from failed
collaborations across different contribution distributions. This allows
us to compare the efficiency of each norm, revealing that C-Norms
generally foster more successful collaborations across a wider range of
scenarios.

Our analyses include robustness checks with varying parameter values for
\(\hat{c}\), different probability distributions of contributions, and
different values of \(\chi, \varepsilon\). These extensions are detailed
in the Supplementary Information.

\printbibliography[title=References]

\appendix

\newpage

{\Huge Supplementary Information}

\newpage

\newcommand{\Jw}{J_w}
\newcommand{\Sw}{S_w}
\newcommand{\Jf}{J_f}
\newcommand{\Sf}{S_f}

\section{Model details}

We are modeling the evolution and social consequence of different norms for author order.  We will be considering two potential norms.

\begin{itemize}
    \item Contribution sensitive norm (C-norm): This is a norm whereby the author who contributed a larger fraction to the final paper is indicated in some way. This might be a norm like ``first author contributed more.''  We will assume that there is only one such norm in play.
    \item Contribution insensitive norm (I-norm): This norm conditions the author order on something exogenous to the contribution on this paper. So norms like ``authors are listed in alphabetical order'' or ``senior authors always listed last.''
\end{itemize}

For simplicity will consider a game between two players Junior and Senior.  We will assume that the I-norm will list them in that order (Junior, Senior). That might be because of the alphabetical order of their names, because Senior is the senior author, or some other reason not determined by relative contribution.

\section{The evolution of norms}

\subsection{Game and equilibria}

To model the evolution of norms we will use the following model: Senior and Junior each decide antecedently on a norm.  If their norms are compatible, then they collaborate; if not, they publish alone.  

What does ``compatible'' mean in this context?  Obviously if they both choose the same norm (both choose the I-norm or the C-norm) then their choices are compatible.  Also, if Senior chooses the I-norm and Junior chooses the C-norm, these choices are compatible.  In this case, neither is insisting on being first, and so we presume they flip a coin to decide which norm to use.

Nature decides how important are the contributions of each Junior and Senior.  Let the value of Junior's contribution be $v_j$ and Senior's contribution be $v_s$.  We don't really care much about the total value here, and so we will normalize it to 1.  Henceforth we will talk about Junior's contribution as $c_j = v_j/(v_j + v_s)$.

If Junior and Senior have already decided to collaborate, they produce a paper that has value $1 + \hat{c}$.  That value accounts for the normalized value of each of their contributions and some added value that is had by collaborating ($\hat{c}$). If they decided not to collaborate, they publish alone. In which case, Junior produces a paper that has value $c_j$ and Senior produces a paper that has value $1-c_j$.

\paragraph*{Aside:} {\itshape It is tempting to think of the proportion of contribution as the amount of work measured in hours (or whatever). But, we don't want to impose such an interpretation. At its most fundamental it is simply the proportion of credit that a fully informed scientific community would give to each author. This credit might reflect temporal investment, intellectual novelty, methodological sophistication, interpretive insight, or some amalgamation of factors that vary considerably across disciplines. In experimental fields, credit might skew toward those who designed protocols or collected data; in theoretical work, toward those who conceived frameworks or derived proofs; in computational research, toward those who architected algorithms or curated datasets. The scientific community's calculus remains opaque and contestable, but the existence of meaningful gradations in perceived contribution is undeniable.} $\Box$

~\\

We don't care much about the structure of the distribution of work.  But we will need three quantities to be well defined.  One is the probability that Junior did more work, given by $w_j \in (0,1)$.  The second is the conditional expectation, {\it given} that Junior did more work, how much do we expect they contributed. This is given by $b_j \in (0.5,1)$.  Similarly for the expectation for for how much work Senior did, if Senior did more work $b_s \in (0.5, 1)$.

Lastly we need the unconditional expectation for the distribution.  This is $\mu_j = w_j b_j + (1-w_j)(1-b_s)$ and $\mu_s = 1 - \mu_j$

\paragraph*{Aside:} {\itshape The order of operations matters here. They don't collaborate if they don't agree on a norm.  If we did it in the other order, say if they collaborated first and nature decides who made the greater contribution, then if Junior contributed more they would {\it always} publish since the norm disagreement wouldn't matter. This requires a much more complicated game because now Junior and Senior can adopt strategies like ``when I contribute more, demand the C-norm, but when I do less, demand the I-norm.''  In fact, it could be even more complicated, any function from $c_j \to \{$C-norm, I-norm$\}$ could be a potential strategy. Because of this complication, this version of the game will be left to future work. $\Box$}

~\\

We treat the scientific community as an unmodeled third player who knows Senior's and Junior's mixed strategies, knows the underlying probability distribution of contribution, and observes the published author order.  They use that information to calculate the expected amount of work that each did on this paper and assign that as the ``credit'' each receives.  

At this stage, we model the community as making perfectly unbiased credit allocations. After establishing some baseline results with this model, we will subsequently add in first author bias and Matthew Effect bias, as discussed in the main text ($\varepsilon, \chi$).

Given the structure of the game, if the third party observes Senior listed first, this means that Senior definitely did more work. So anytime Senior is listed first, the payoffs are $1-b_s$ for Junior (the expected amount of work that Junior did {\it given} that Junior did less than Senior) and $b_s$ for Senior (the expected amount of work that Senior did, conditional on having done more than Junior).

However, if Junior is listed first the probability that Junior did more work is more complicated.

Let's name several relevant events:
\begin{itemize}
    \item $\Jf, \Sf$: Junior/Senior is listed first on a published paper
    \item $C$: They successfully collaborate
    \item $\Jw, \Sw$: Junior/Senior did more work on the published paper
\end{itemize}

Also, we will represent the probability that Junior plays ``I-norm'' to be $p_j$ and the probability that Senior plays ``I-norm'' to be $p_s$.

First, we calculate $P(C)$, the probability that they successfully agree to collaborate.  This requires that they have compatible norms.  The only incompatible norm is (I-norm, C-norm), this is where Junior demands to be first regardless and Senior wants to go first if they do more work. So $P(C) = 1 - p_j(1-p_s)$.

Whether Junior or Senior does more work on the paper is independent of whether they agree to collaborate so, critically $P(\Jw) = w_j$ and $P(\Jw|C) = w_j$ (and similarly for $P(\Sw) = P(\Sw|C) = 1-w_j$).

What our third party wants to know is, given that they observe a successful collaboration with Junior listed as first author, how likely is it that Junior did more work?  Namely they want to know what is $P(\Jw | \Jf \& C)$?

\begin{equation*}
    P( \Jw | \Jf \& C) = \frac{P(\Jf \& C | \Jw)P(\Jw)}{P(\Jf \& C | \Jw)P(\Jw) + P(\Jf \& C | \Sw)P(\Sw)}
\end{equation*}

Because of the independencies listed above, $P(\Jf \& C | \Jw) = P(C)$.  That is, if Junior did more work, the probability they are listed first is just the probability that they successfully collaborate.

\begin{equation*}
    P( \Jw | \Jf \& C) = \frac{P(C)P(\Jw)}{P(C)P(\Jw) + P(\Jf \& C | \Sw)P(\Sw)}
\end{equation*}

$P(\Jf \& C | \Sw)$ is the probability that despite Senior doing more work, Junior is listed first on a published paper.  This occurs when they agree on (I-norm, I-norm) or half of the time when they say (C-norm, I-norm).  So, $P(\Jf \& C | \Sw) = p_j p_s + 0.5(1-p_j)p_s$.

Substituting all this in gives us:
\begin{equation*}
    P( \Jw | \Jf \& C) = \frac{ (1 - p_j(1-p_s))w_j}{(1 - p_j (1-p_s))w_j + (p_j p_s + 0.5(1-p_j)p_s)(1-w_j)}
\end{equation*}
We will call this $m_j$ for simplicity.  (When $p_j=1$ and $p_s=0$, this is undefined because the condition has measure 0.  For simplicity, we will say that $m_j = 0$ in this case.)

When Junior is listed first, his payoff will be:
\begin{equation*}
    m_j b_j + (1-m_j)(1-b_s)
\end{equation*}
And when Junior is listed first the payoff to Senior is:
\begin{equation*}
    m_j (1-b_j) + (1-m_j) b_s
\end{equation*}

When Senior is listed first, it is certain they did more work.  There is no condition under which they did less work and was listed first. So, the payoffs will be $b_s$ for Senior and $(1-b_s)$ for Junior.

Now we need to know what is the ex ante probability that they will collaborate and Junior or Senior will be listed first ($f_j$ or $f_s$ respectively):
\begin{align*}
    f_j & = w_j(1-p_j(1-p_s)) + (1-w_j)(p_jp_s + (1-p_j)p_s/2) \\
    f_s & = (1-w_j)((1-p_j)(1-p_s) + (1-p_j)p_s/2)
\end{align*}

So, now we are in a position to give the payoff functions for the two players:
\begin{align*}
\pi_j(p_j, p_s) = &  f_j \big(m_j b_j + (1-m_j)(1-b_s)\big)(1+\hat{c}) + \\
                & f_s(1-b_s)(1+\hat{c}) + p_j (1 - p_s)\mu_j \\
\pi_s(p_s, p_j) = &  f_j \big(m_j (1-b_j) + (1-m_j) b_s \big)(1+\hat{c}) + \\
                & f_s b_s(1+\hat{c}) + p_j (1 - p_s)\mu_s
\end{align*}

While this game initially appears quite unwieldy, it turns out many of the larger expressions cancel out and you end up with a relatively simple game. 

\subsection{Comparative statics}

It is worth noting that so long as $\hat{c}>0$ and $0 < p_j, p_s < 1$, then the derivative for $\pi_j$ with respect to $p_j$ is always negative and the derivative for $\pi_s$ with respect to $p_s$ is always positive.  This means that both agents would prefer to alter their preferred norm in order to increase the probability that successful collaboration occurs (see Mathematica notebook).

Second, when $p_j = 0$, the derivative of $\pi_s$ with respect to $p_s$ is zero as is $\pi_j$ when $p_s = 1$. This means that when Junior is fully committed to the C-norm, Senior is indifferent between all of their strategies. And when Senior is fully committed to the I-norm, then Junior is indifferent between all strategies.  These are cases where collaboration is guaranteed.  Because of the rationality of the scientific community, the indifferent player cannot improve their expected credit by altering their strategy. 

Instead of considering all these equilibria, we will focus on the two pure strategy equilibria. There are two ways we can evaluate preferences between equilibria: ex ante (before they know relative contributions) and ex post (after they know relative contributions).

First, ex ante. In this situation each author doesn't know whether they will contribute more and must decide which equilibrium they prefer.  In the I-norm equilibrium, both players are paid their expected contribution, $\mu_j$ and $\mu_s$ respectively. 

In the C-norm equilibrium, note that Junior expected payoff is $\mu_jb_j + (1-\mu_j)(1-b_s) = \mu_j$. Senior similarly is: $\mu_j(1-b_j) + (1-\mu_j)b_s = \mu_s$.  So their expected payoff under the C-norm is the same.

Ex post, of course, things are different.  The payoff in the I-norm is the same.  But now the person who did more work would prefer to be in the equilibrium with the C-norm, because their payoff would be higher.  The other player prefers the I-norm.

\subsection{Replicator dynamics}

Now let's consider evolution in this game with the replicator dynamics. Because of what was already shown, all points where either (\emph{a}) $p_j = 0$ and $p_s \in [0,1]$ or (\emph{b}) $p_j \in [0,1]$ and $p_s=1$ are fixed points.  In addition $p_j = 0$ and $p_s = 0$ is a fixed point, but because it is not a Nash equilibrium, it is unstable.

This result entails that many mixed norms are possible.  

\subsection{Adding in first author bias}

When deciding how much credit to give to each author, the ``correct'' result from the game described above is now averaged with an accounting method that assigns all the credit for the entire paper $(1 + \hat{c})$ to the first author. This is meant to represent cognitive biases or constraints whereby the first author is the only one remembered.  We will represent the probability of the ``stupid'' strategy with $\varepsilon$.

\begin{align*}
    \pi_j(p_j, p_s) = & f_j \big((1-\varepsilon)(m_j b_j + (1-m_j)(1-b_s)) + \varepsilon\big)(1+\hat{c}) +\\
    & f_s(1-\varepsilon)(1-b_s)(1+\hat{c}) + p_j (1 - p_s)\mu_j \\
    \pi_s(p_s, p_j) = & f_j \big((1-\varepsilon)(m_j (1-b_j) + (1-m_j) b_s)\big)(1+\hat{c}) + \\
    & f_s\big((1-\varepsilon)b_s + \varepsilon\big)(1+\hat{c}) + p_j (1 - p_s)\mu_s
\end{align*}

This generalization of the game entails the old version as a special case (where $\varepsilon = 0$).  However, the equilibria change significantly with the introduction of $\varepsilon > 0$. For the small values of epsilon that interest us, all but one of the mixed strategy equilibria are eliminated.

\paragraph*{Aside:} {\itshape It is worth noting some interesting properties when $\varepsilon$ is large.  In that case, the I-norm is no longer stable, as Senior would rather not collaborate and publish on their own rather than seeding the lion's share of the credit to Junior.

Whether the C-norm is stable now depends on Junior.  If $w_j$ is sufficiently high (relative to $\varepsilon$ and $\hat{c}$) that Junior will get almost all the credit often enough, then the C-norm is the unique stable equilibrium.  On the other hand if $w_j$ is small, then Junior would prefer to publish alone and the only stable equilibrium is where $p_j=1$ and $p_s=0$---where they never collaborate. 

While formally interesting, this is not our focus for this paper. Our second model will more accurately model the incentives for failed collaboration. Instead, we will focus on cases where the non-collaboration state is not an equilibrium. In these cases there is either one stable equilibrium, which is the C-norm, or there are two stable equilibria, which are the C- and I-norms.  When there are two stable equilibria there is a third unstable mixed equilibrium. $\Box$}

~\\

We will now look at a series of examples where the values for $w_j$, $b_j$, and $b_s$ are taken from a Beta distribution with parameters $\alpha$ and $\beta$.  (It is easy to calculate $w_j = \alpha/(\alpha + \beta)$.  $b_j$ and $b_s$ are calculated numerically.

\begin{figure}
    \centering
    \includegraphics[width=\textwidth]{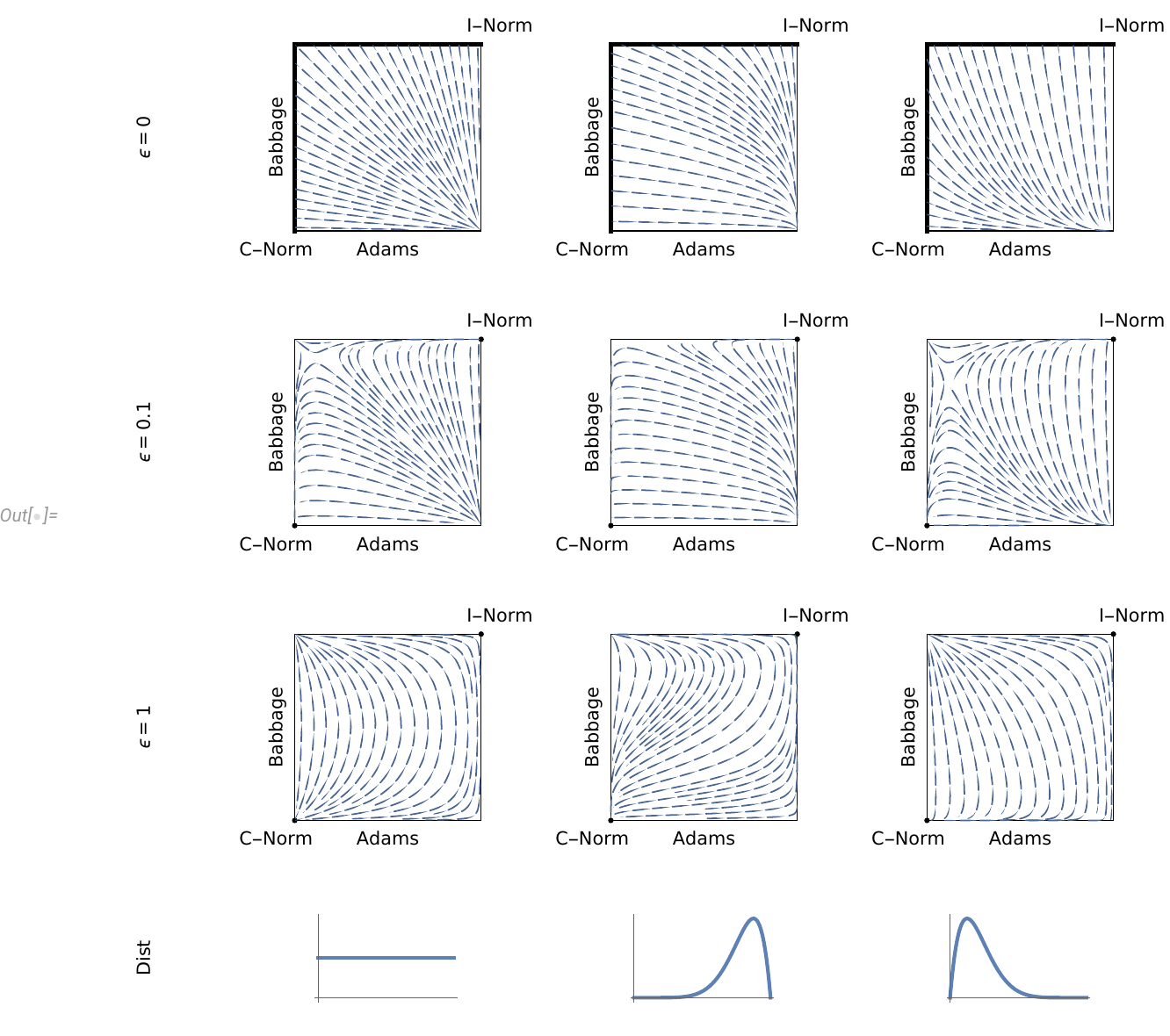}
    \caption{Illustrations of the replicator dynamics in nine different cases.  The rows represent three different values of $\varepsilon$ ($\varepsilon=0, 0.1, 1$).  The columns each represent a different beta distribution prior.  The left is a uniform distribution, the center is a a prior which heavily weights Junior doing more work ($\alpha = 8$, $\beta = 2$) and the right heavily weights Senior ($\alpha=2$ and $\beta=8$). $\hat{c} = 1$}
    \label{f:replicator-illustration-n}
\end{figure}

Figure~\ref{f:replicator-illustration-n} gives nine examples of the replicator dynamics in this game where $\hat{c} = 1$

\begin{figure}
    \centering
    \includegraphics[width=\textwidth]{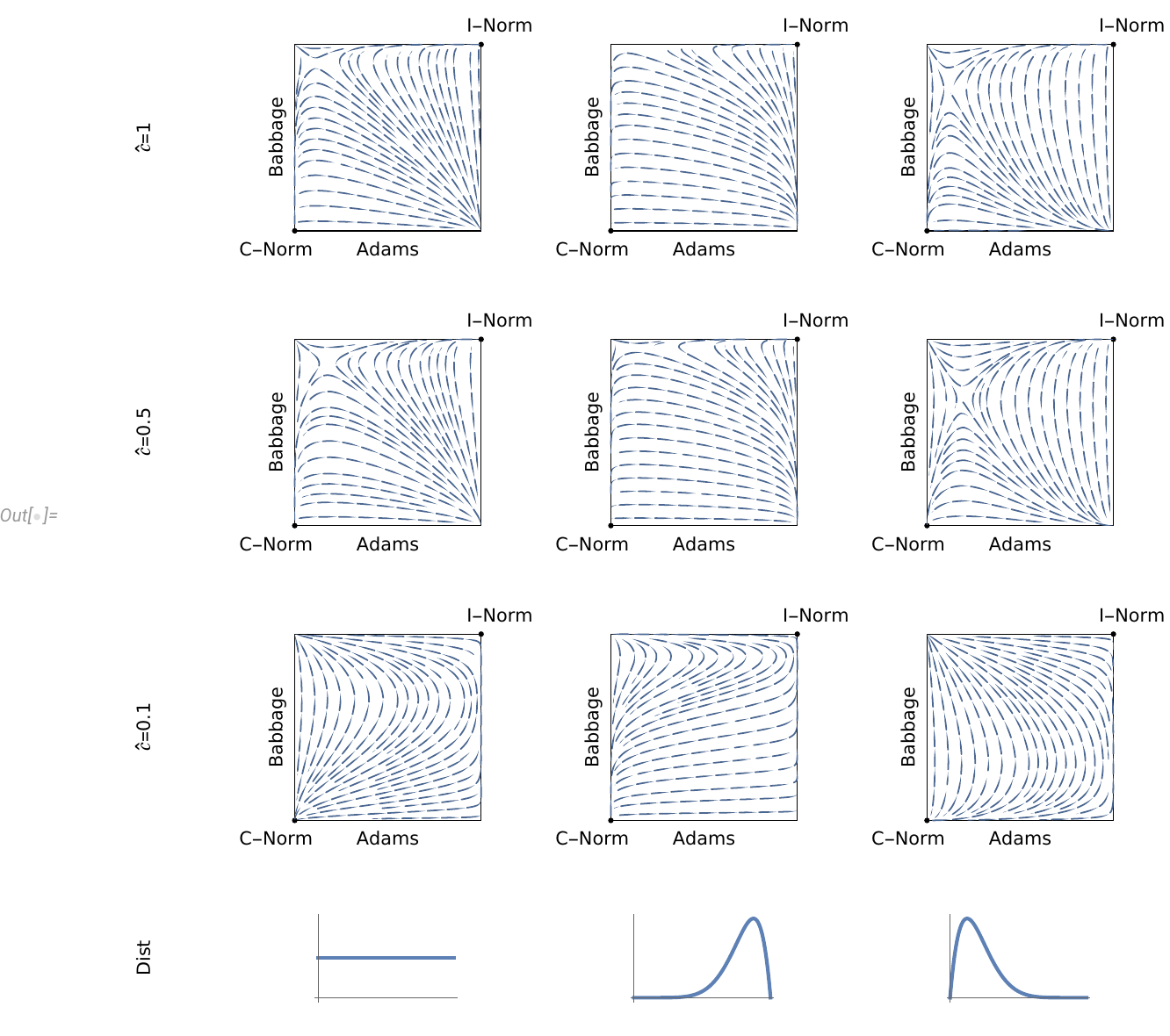}
    \caption{Illustrations of the replicator dynamics in nine different cases.  The rows represent three different values of $\hat{c}$ ($\hat{c}=1, 0.5, 0.1$).  The columns each represent a different beta distribution prior.  The left is a uniform distribution, the center is a a prior which heavily weights Junior doing more work ($\alpha = 8$, $\beta = 2$) and the right heavily weights Senior ($\alpha=2$ and $\beta=8$). $\varepsilon = 0.1$}
    \label{f:replicator-illustration-chat}
\end{figure}

We will mostly focus on cases where $\varepsilon$ is small but not 0. To consider the effects of $\hat{c}$ in such a case, see figure~\ref{f:replicator-illustration-chat}.

\begin{figure}
\centering
    \begin{subfigure}[b]{0.45\textwidth}
        \centering
        \includegraphics[width = \textwidth]{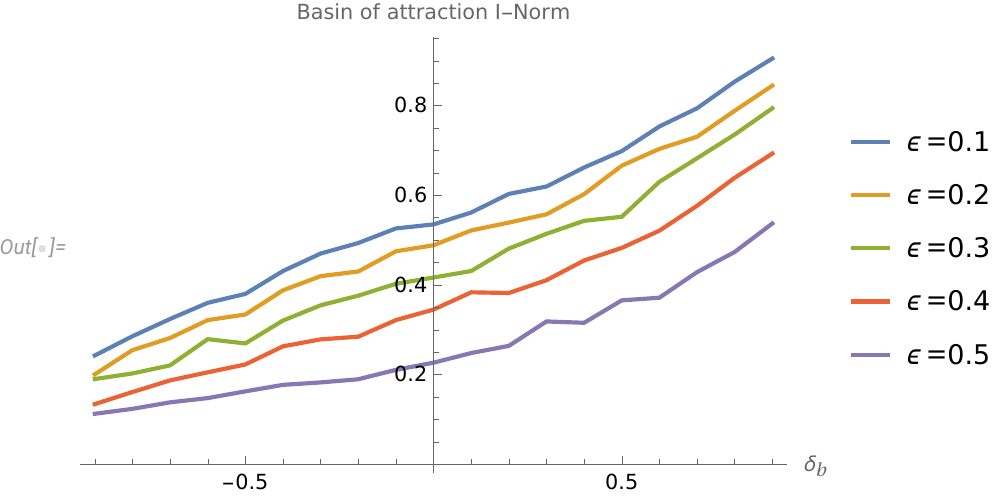}
        \caption{Basin of attraction for I-norm}
    \end{subfigure}
    \hfill
    \begin{subfigure}[b]{0.45\textwidth}
        \centering
        \includegraphics[width = \textwidth]{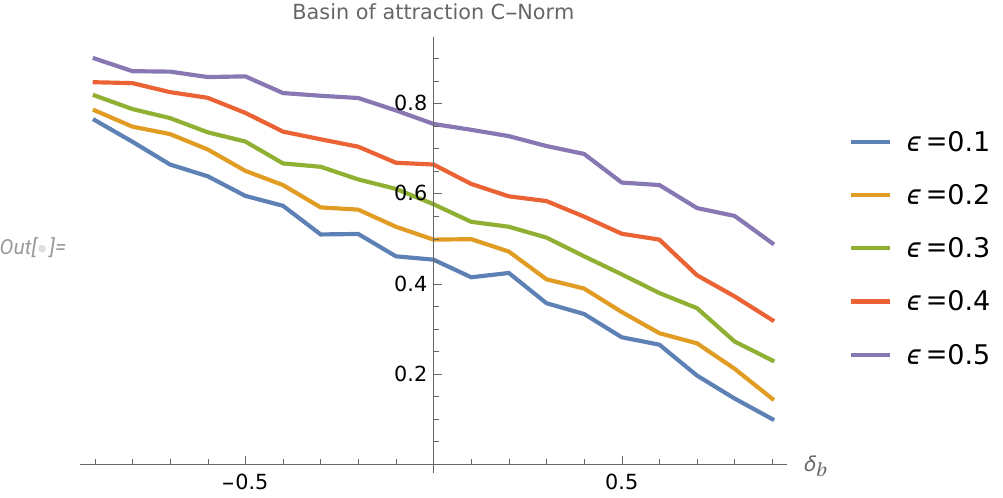}
        \caption{Basin of attraction for C-norm}
    \end{subfigure}
    \caption{Basins of attraction for the two norms. Horizontal axis is $\delta$ which is the difference between the expected contribution of Senior and Junior ($\delta = 1 - 2\mu_j$). Vertical axis is the size of the basin of attraction. Each line represents a different value of $\varepsilon$. $\hat{c}=1$, $\chi = 0.1$}
    \label{f:epsilonsearch}
\end{figure}

Suppose that we take the underlying distribution of contributions to be a Beta distribution with parameters $\alpha = a$ and $\beta = 100-a$.  So, as we increase the parameter $a$ that means Junior is expected to contribute a larger share.  The various parameters as well as the basin of attraction for the I-norm are shown in figure~\ref{f:epsilonsearch}.

The somewhat surprising result is that the larger the expected contribution of Junior, the more likely the pair will end up in the C-norm.  This occurs because of the relative speed of evolution in the two populations. When Junior is expected to contribute more, they stand to lose more from a failed collaboration than does Senior.  So, Junior evolves faster than Senior.  As discussed in the main text, there are various other ways that the same dynamic could arise.

\subsection{Adding in the Matthew Effect}

If we interpret this as a game between a junior and senior where the I-norm is ``senior author last,'' there remains another possible bias to consider: the Matthew Effect.  The Matthew Effect, first described by Merton, is one where the more famous collaborator is given more credit than less famous collaborators. In this context, we will assume the senior author is the more famous one, who will potentially benefit from the Matthew Effect.

So, now we have three potential ways credit might be assigned:
\begin{enumerate}
    \item The ``correct'' way according to the Bayesian calculations about expected credit
    \item According to ``first author bias'' where credit for the whole paper is given to whoever is listed first
    \item According to the Matthew Effect where credit for the whole paper is given to Senior (regardless of author order)
\end{enumerate}

The probability for each of these three will be given by: $(1-\varepsilon)(1-\chi)/(1 - \varepsilon\chi)$, $\varepsilon(1-\chi)/(1 - \varepsilon\chi)$, and $(1-\varepsilon)\chi/(1 - \varepsilon\chi)$.  We assume that $\varepsilon > 0$, $\chi > 0$, and $\varepsilon + \chi < 1$.

To understand the effect of this bias, let us first consider the extreme version where $\chi = 1$ (and $\varepsilon = 0$). Under this parameter setting Senior gets all the credit when they publish together. If Senior ever employs the C-norm, Junior will do better by demanding the I-norm in order to get their collaboration to fail. But in response, Senior can just demand the I-norm as well and get all the credit. If Senior demands the I-norm, it doesn't matter what Junior demands because they will collaborate and Senior will get all the credit.  So there are an infinity of equilibria where $p_s = 1$ and $p_j \in [0,1]$. 

For large $\chi < 1$ and sufficiently small $\hat{c}$ this dynamic remains.  However, if $\hat{c}$ is large enough collaboration is beneficial for Junior.  Now in response to Senior demanding the C-norm, Junior also wants to demand the C-norm.  However, the pressure on Senior to secure collaboration is much stronger than the pressure on Junior. As a result, the C-norm equilibrium remains unstable. Once $\chi$ is small enough or $\hat{c}$ is large enough, the C-norm becomes stable.

Like with $\varepsilon$, our main interest is not in the dynamics for large values of $\chi$, but rather for small values.

For small values of $\chi$ it creates stronger selection pressure for Senior to collaborate and thus increases the basins of attraction for the I-norm (much like the case where Senior is expected to make a larger contribution).

\begin{figure}
\centering
    \begin{subfigure}[b]{0.45\textwidth}
        \centering
        \includegraphics[width=\textwidth]{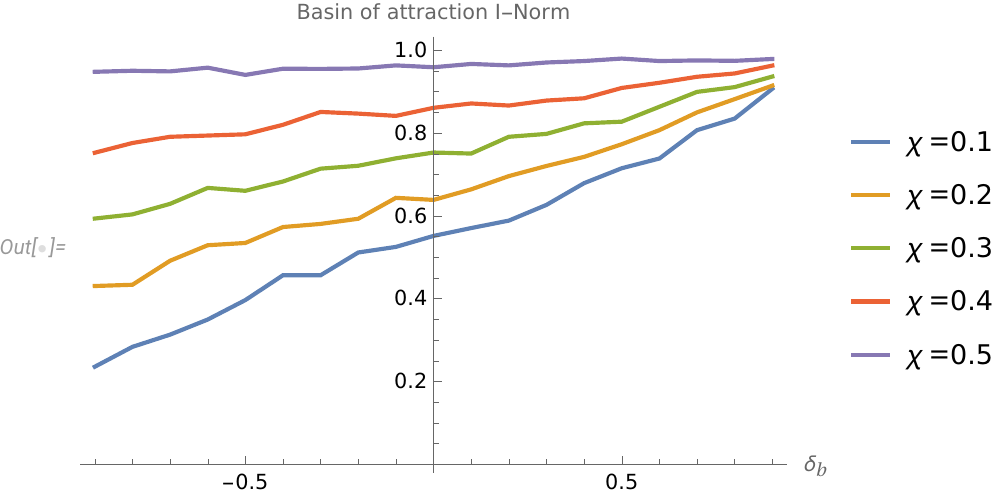}
        \caption{Basin of attraction for I-norm}
    \end{subfigure}
    \hfill
    \begin{subfigure}[b]{0.45\textwidth}
        \centering
        \includegraphics[width=\textwidth]{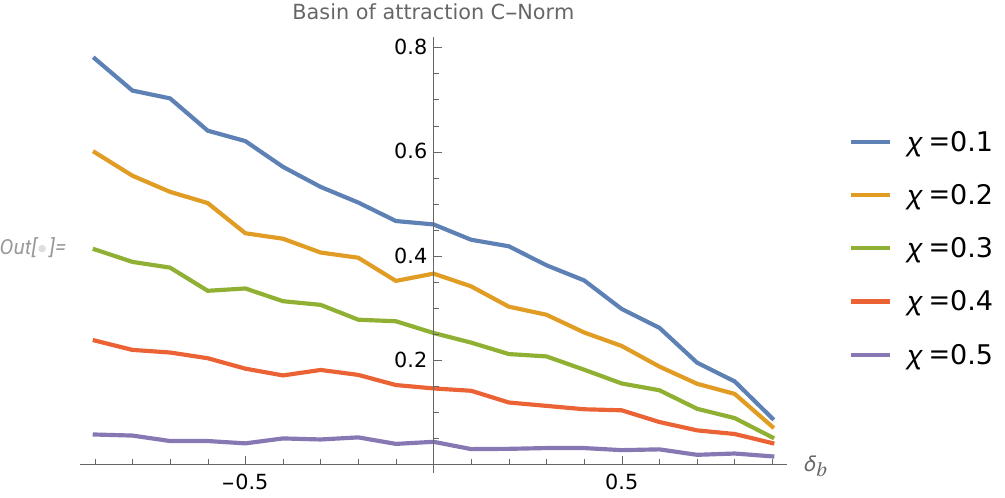}
        \caption{Basin of attraction for C-norm}
    \end{subfigure}
    \caption{Basins of attraction for the two norms. x-axis is $\delta$ which is the difference between the expected contribution of Senior and Junior ($\delta_b = 1 - 2\mu_j$). y-axis is the size of the basin of attraction. Each line represents a different value of $\chi$. $\varepsilon = 0.1$ and $\hat{c} =1$.}
    \label{f:chisearch}
\end{figure}

Figure~\ref{f:chisearch} shows how $\chi$ has a similar effect on the basin of attraction as increasing Senior's expected contribution.  Both work for essentially the same reason.

\section{Will they collaborate?}

In the previous version of the model, we assumed that if they agreed to collaborate they would follow through after learning the value of their respective contributions.  But, what if they maintain an option of ``taking their toys and leaving?'' That is, they can take the part of the project that they did and publish it independently. 

For this game, we will assume that exactly one norm is in place and this is known by everyone.  So either the I-norm or the C-norm is the norm.  Players learn their relative contributions, call that $c_j$ for Junior and $1-c_j$ to Senior.  

The players choices are either ``continue collaboration'' or ``go it alone.'' This decision can be made unilaterally, and if one party decides to do it, they both will publish independently.  Each player's strategy is then a function from $c_j$ to the set $\{$collaborate, go alone$\}$.

As  above, we are normalizing the value for the two single authored papers to 1, so when they go it alone the payoffs are $c_j$ for Junior and $(1-c_j)$ for Senior.  When they collaborate, the value of the joint paper is $(1+\hat{c})$, and this is divided between them according to the underlying distribution and the norm.

We will assume that credit is assigned by the ``correct'' payoff function above.  I.e. $\varepsilon = \chi = 0$.  Addition of these biases would shift the results, but so long as they are small, would not yield qualitatively different conclusions.

\subsection{I-norm}

Suppose that the I-norm is in place. This means that Junior will always be listed first, regardless of how much he/she contributes. Since this is the norm in place, the third party has no idea who contributed more to the paper.  As a result, when they collaborate Junior will always receive payoff $\mu_j(1+\hat{c})$ and Senior will receive payoff $(1-\mu_j)(1+\hat{c})$.

The value of going it alone for Junior is $c_j$, so if that is greater than $\mu_j(1+\hat{c})$ they will refuse to collaborate.  Notice that this won't happen when $c_j$ is near $\mu_j$. It only happens when the draw of $c_j$ is significantly above its expectation

For Senior, the value of publishing independently is $(1-c_j)$, so they will refuse collaboration if this exceeds their expected payoff from joint publication, $(1-\mu_j)(1+\hat{c})$. As with Junior, this occurs when Senior's actual contribution substantially exceeds their ex ante expected contribution: Senior will choose independent publication if and only if their realized work on the project significantly surpasses the amount the scientific community anticipated they would contribute.

\begin{figure}

    \centering
    \includegraphics[width=0.5\textwidth]{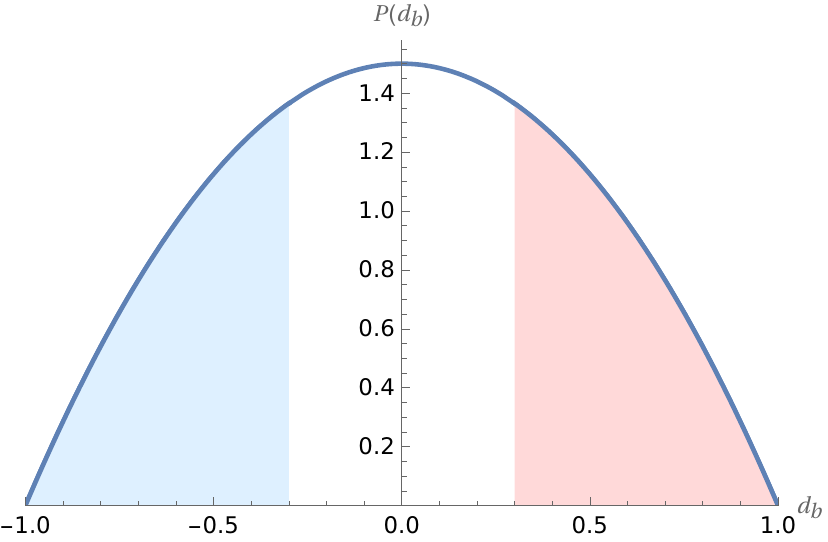}
    \caption{Collaboration failures in the context of the I-norm. $\hat{c} = 0.3$.  The underlying distribution is a beta distribution with $\alpha = \beta = 2$. The red region is where Senior refuses to collaborate, and the blue region is where Junior refuses to collaborate. The x-axis is the value of $d = 1 - 2c_j$, the value of Senior's contribution minus Junior contribution.}
    \label{f:INormFailure}
\end{figure}

In summary, under the I-norm, collaborations fail when either author's contribution substantially exceeds expectations---a pattern we term ``main contributor resentment.''\footnote{This terminology serves as a mnemonic device; the model contains no actual resentment since collaboration is preemptively avoided. Rather, the label captures the intuition that the primary contributor would feel undercompensated relative to their actual work if collaboration proceeded.} This is illustrated in figure~\ref{f:INormFailure}.

\begin{figure}
    \centering
    \includegraphics[width=\textwidth]{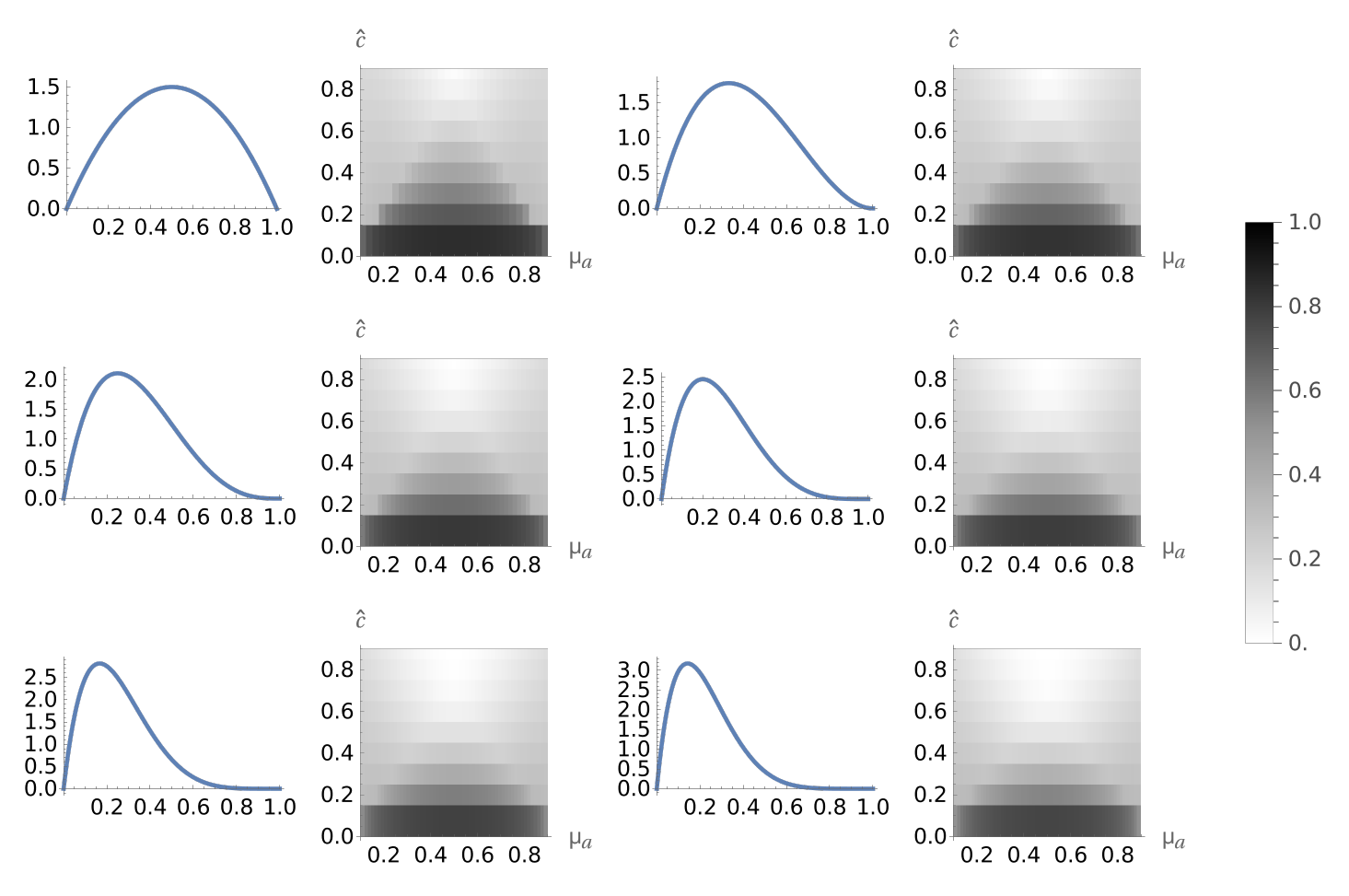}
    \caption{Heatmap indicating where collaborations fail under the I-norm. Next to each heatmap is an illustrative Beta distribution showing the shape of the distribution (the mean will move around depending on the relevant point in the heatmap).  White indicates low ex ante probability of failure, black  indicates high probability of failure.  The y-axis is $\hat{c}$, the x-axis is $\mu_j$. These are plots of ex ante probability, meaning that it's the probability of failure before the agents know their relative contributions.}
    \label{f:nonbayesian-inorm-failrate}
\end{figure}

We can also calculate ex ante how likely collaborations are to fail.  For example, when $\hat{c} = 0$ (as an extreme), then collaboration will fail with probability 1.  One party will have contributed more than the credit they will receive, and so would prefer to go it alone.  This will also often hold for small $\hat{c}$.  As $\hat{c}$ gets larger, collaboration will happen more often, but can still fail for moderate values.  (In the limit $\hat{c}$ approaches $\infty$, then it will never fail.)  Figure~\ref{f:nonbayesian-inorm-failrate} illustrates how often collaborations fail under various parameters.

\begin{figure}
    \centering
    \includegraphics[width=\textwidth]{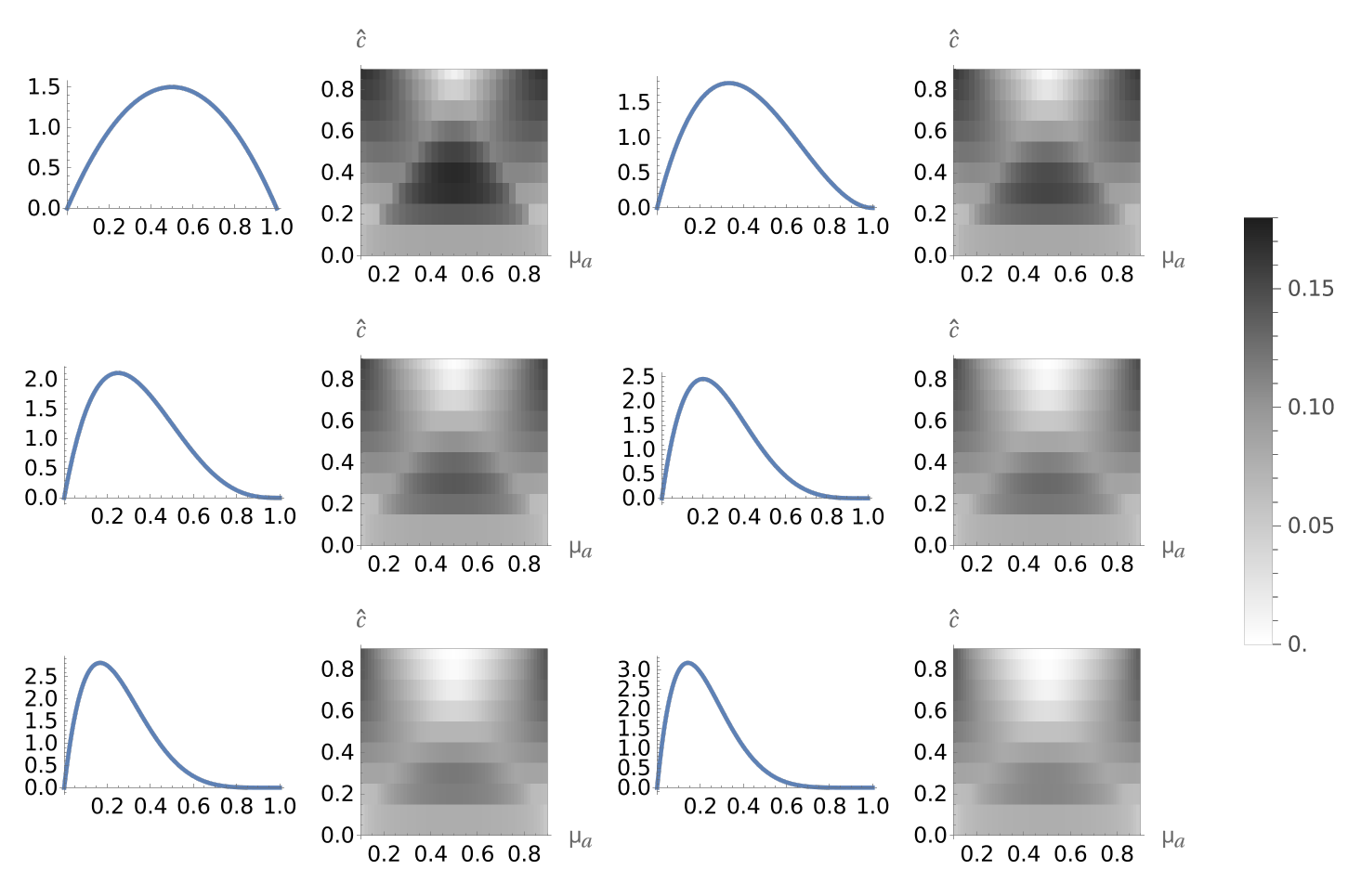}
    \caption{Heatmap indicating how much of the public good is lost given under the I-norm. Next to each heatmap is an illustrative Beta distribution showing the shape of the distribution (the mean will move around depending on the relevant point in the heatmap).  Black indicates high ex ante cost, white indicates low.  The y-axis is $\hat{c}$, the x-axis is $\mu_j$.}
    \label{f:nonbayesian-inorm-publicgood}
\end{figure}

The additional benefit from collaboration, $\hat{c}$, is a public good. Beyond looking at when collaborations fail, we can also ask how much is lost from their failure.  When $\hat{c} = 0$, collaboration always fails, but also there is no harm from it.  However, when $\hat{c}$ is larger, then it is worse.  Figure~\ref{f:nonbayesian-inorm-publicgood} illustrates the ex ante cost expected from failures to collaborate.

\subsection{C-norm}

The C-norm analysis requires examining two distinct cases: when Junior contributes more and when Senior contributes more.

\textbf{Case 1: Junior contributes more} ($c_j > 0.5$). Junior will be listed first and receive credit $b_j$. They will refuse collaboration if $c_j > b_j(1+\hat{c})$. This represents a more stringent condition than the analogous failure under the I-norm, since $c_j$ must substantially exceed the conditional expectation $b_j$ (Junior's expected contribution given that they contributed more than Senior).

Under Case 1, Senior may also refuse if $(1-c_j) > (1 - b_j)(1+\hat{c})$. This occurs when Senior's contribution exceeds what they would typically expect to contribute when performing less work than Junior. In essence, the C-norm discourages collaborations where contributions are approximately equal, since the second-listed author receives disproportionately less credit despite similar effort.

\textbf{Case 2: Senior contributes more} ($c_j < 0.5$). Senior will refuse collaboration if their contribution is extraordinarily large, $(1-c_j) > b_s(1+\hat{c})$. Junior will refuse if their contribution nearly matches Senior's despite being slightly less, $c_j > (1 - b_s)(1+\hat{c})$.

The dynamics mirror Case 1, confirming that collaboration remains discouraged for highly unequal contributions. However, unlike the I-norm, the C-norm also discourages collaborations where contributions are nearly equal, since the fixed ordering creates artificial credit distinctions despite similar effort levels.

\begin{figure}
    \centering
    \includegraphics[width=0.5\textwidth]{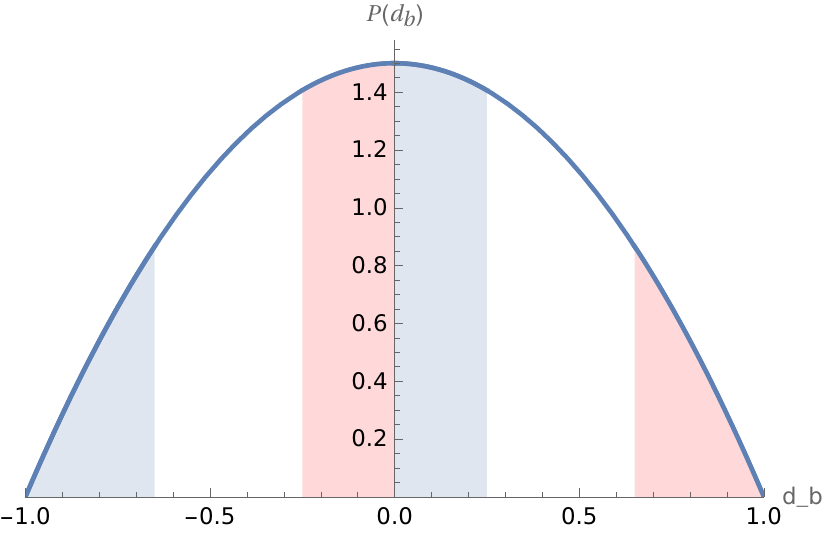}
    \caption{Collaboration failures in the context of the C-norm. $\hat{c} = 0.15$.  The underlying distribution is a beta distribution with $\alpha = \beta = 2$. The red region is where Senior refuses to collaborate, and the blue region is where Junior refuses to collaborate.  The x-axis is the value of $d = 1 - 2c_j$, the value of Senior's contribution minus Junior contribution.}
    \label{f:CNormFailure}
\end{figure}

All this can be seen in figure~\ref{f:CNormFailure}. This illustrates how the C-norm discourages both near-equal contribution collaborations and extremely unequal contribution collaborations.  

\begin{figure}
    \centering
    \includegraphics[width=\textwidth]{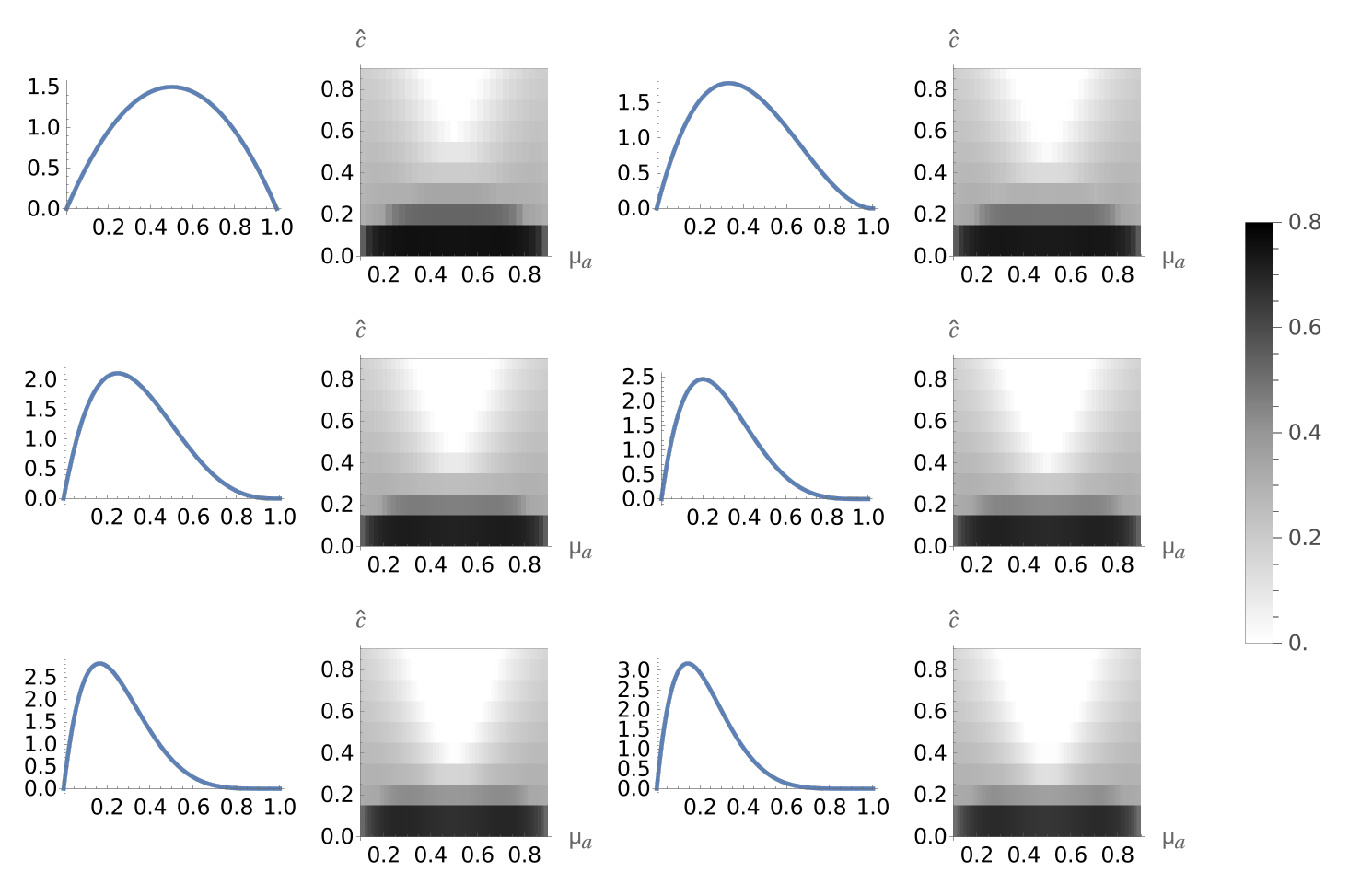}
    \caption{Heatmap indicating where collaborations are likely to fail under the C-norm. Next to each heatmap is an illustrative Beta distribution showing the shape of the distribution (the mean will move around depending on the relevant point in the heatmap).  Black indicates high ex ante probability of failure, white indicates low.  The y-axis is $\hat{c}$, the x-axis is $\mu_j$. These are plots of ex ante probability, meaning that it's the probability of failure before the agents know their relative contributions.}
    \label{f:nonbayesian-cnorm-failrate}
\end{figure}

\begin{figure}
    \centering
    \includegraphics[width=\textwidth]{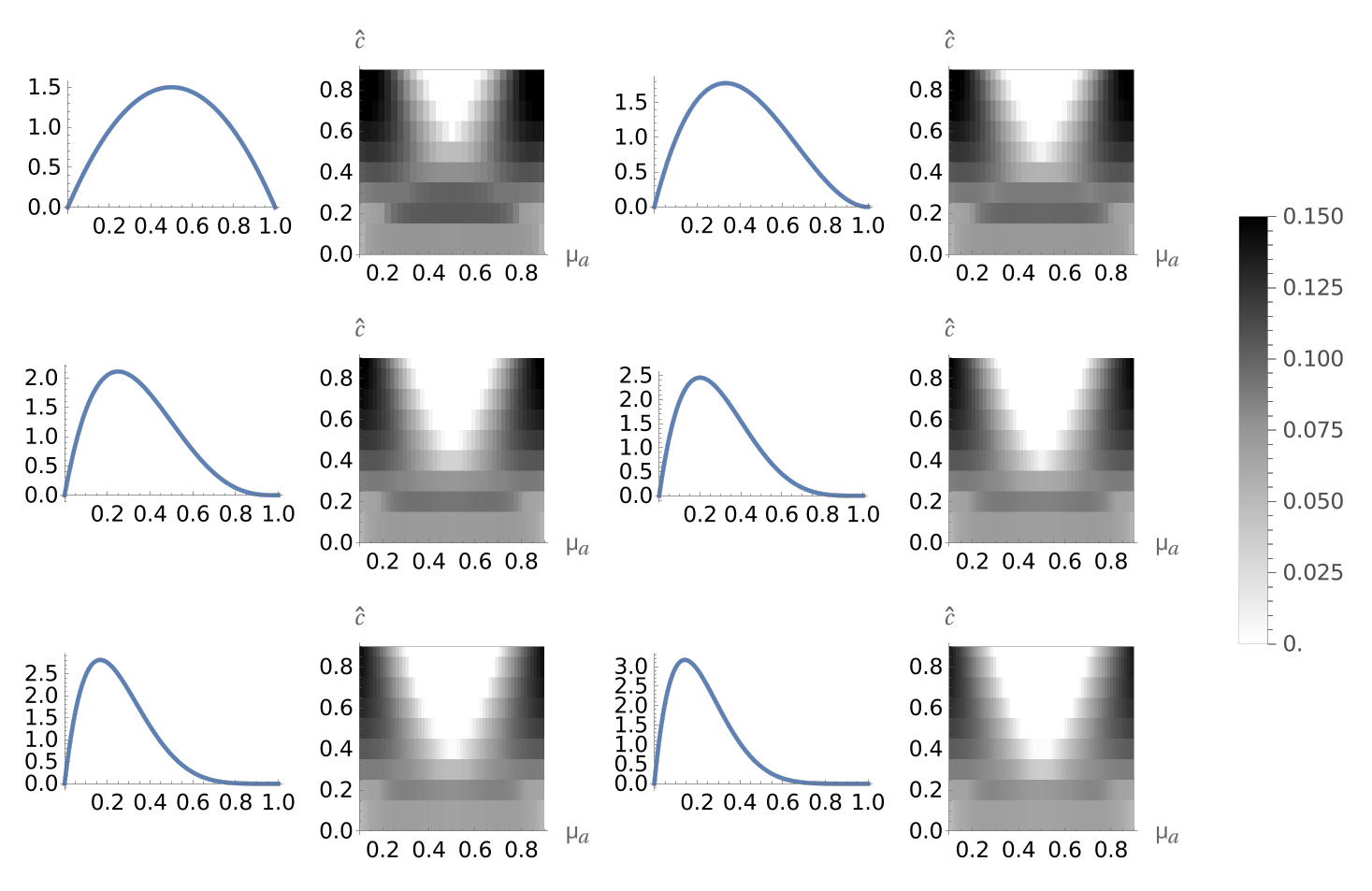}
    \caption{Heatmap indicating how much of the public good is lost under the C-norm. Next to each heatmap is an illustrative Beta distribution showing the shape of the distribution (the mean will move around depending on the relevant point in the heatmap).  Black indicates high ex ante loss of public good, white indicates low.  The y-axis is $\hat{c}$, the x-axis is $\mu_j$. These are plots of ex ante probability, meaning that it's the probability of failure before the agents know their relative contributions.}
    \label{f:nonbayesian-cnorm-publicgood}
\end{figure}

We can now do all the same analysis done with the I-norm. This can be found in figures~\ref{f:nonbayesian-cnorm-failrate} and \ref{f:nonbayesian-cnorm-publicgood}.

\subsection{Failed collaboration under the two norms}

We now examine how the two norms compare. Is one systematically better than the other?

First, consider when collaborations succeed or fail. Under both norms, high collaboration benefits ($\hat{c}$) lead to nearly universal success, while very low benefits result in frequent failure.

Under the I-norm, collaborations succeed when contributions align with expectations. Specifically, collaborations where Junior and Senior contribute close to $\mu_j$ and $(1-\mu_j)$ respectively will proceed. Collaborations fail when either party contributes significantly more than expected. This occurs because the I-norm rewards each author as if they contributed the expected amount. When actual contributions match expectations, the rewards justify collaboration. When they diverge substantially, the undercompensated party prefers independent publication.

Under the C-norm, collaborations at highly unexpected contribution levels also fail. Additionally, the C-norm discourages collaborations with nearly equal contributions. The first author receives credit based on the expected contribution, conditional on contributing more. If their actual contribution falls short of this expectation (approaching 0.5), the second-listed author may prefer independent publication over the diminished credit from second authorship.

Because of this, neither norm dominates the other. Depending on the pattern of contributions, some collaborations might fail under one but succeed under the other.  (Although some collaborations that involve highly unequal contributions will fail under both.)

We now turn to an ex ante analysis: on average, which norm performs better?

\begin{figure}
    \centering
   \includegraphics[width=\textwidth]{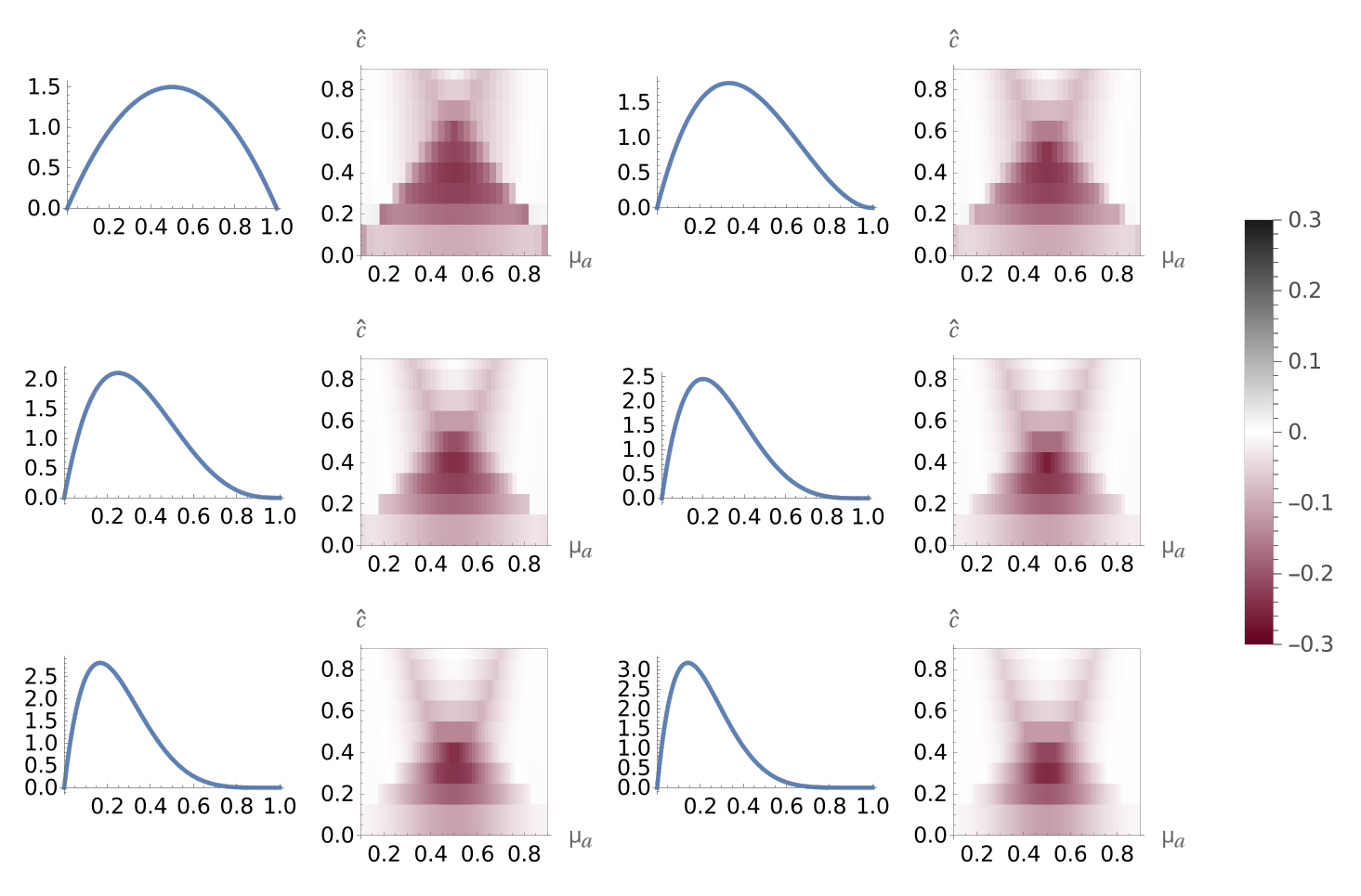}
    \caption{Heatmaps comparing the two norms to one another.  The values compared are the probability of a failed collaboration. Positive values indicate that the I-norm is better than the C-norm (these are colored gray).  Negative values indicate that the C-norm is better than the I-norm (these are colored red).  Zero (white) means they are equivalent.}

    \label{f:nonbayesian-comparison-collab}
\end{figure}

\begin{figure}
    \centering
    \includegraphics[width=\textwidth]{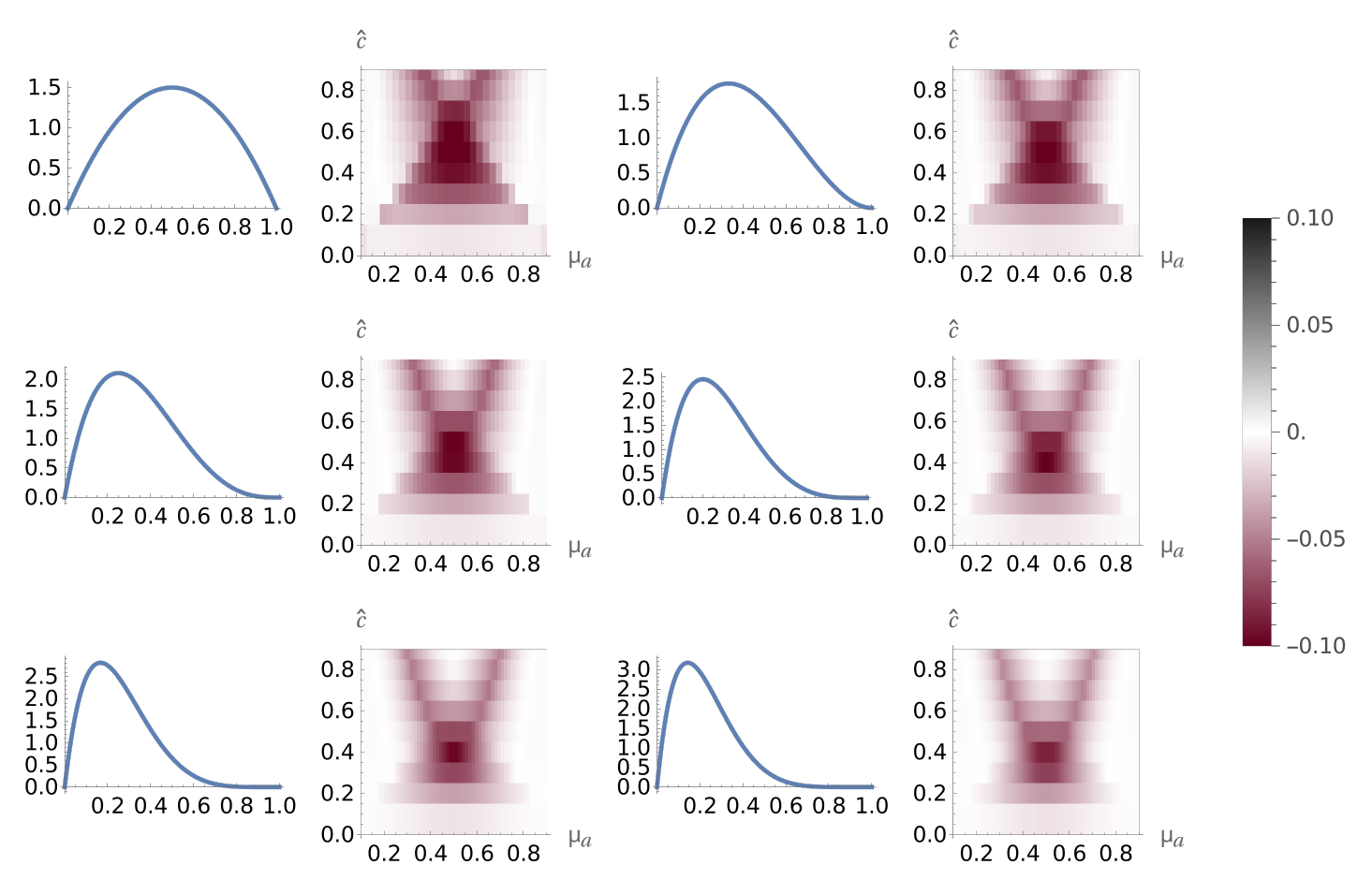}
    \caption{Heatmaps comparing public goods lost under the two norms. Positive values indicate that the I-norm is better than the C-norm (these are colored gray).  Negative values indicate that the C-norm is better than the I-norm (these are colored red).  Zero means they are equivalent.}

    \label{f:nonbayesian-comparison-publicgoods}
\end{figure}

The I-norm generally yields inferior outcomes in ex ante comparisons. However, at extreme values of $\mu_j$, the C-norm can perform marginally worse. This pattern is illustrated in figures~\ref{f:nonbayesian-comparison-collab} and~\ref{f:nonbayesian-comparison-publicgoods}, where negative values indicate superior performance by the C-norm.

\subsection{Equilibrium preference}

Given the possibility of collaboration failure, we now might ask again about preference for the different equilibria.  

\begin{figure}
    \centering
    \includegraphics[width=\textwidth]{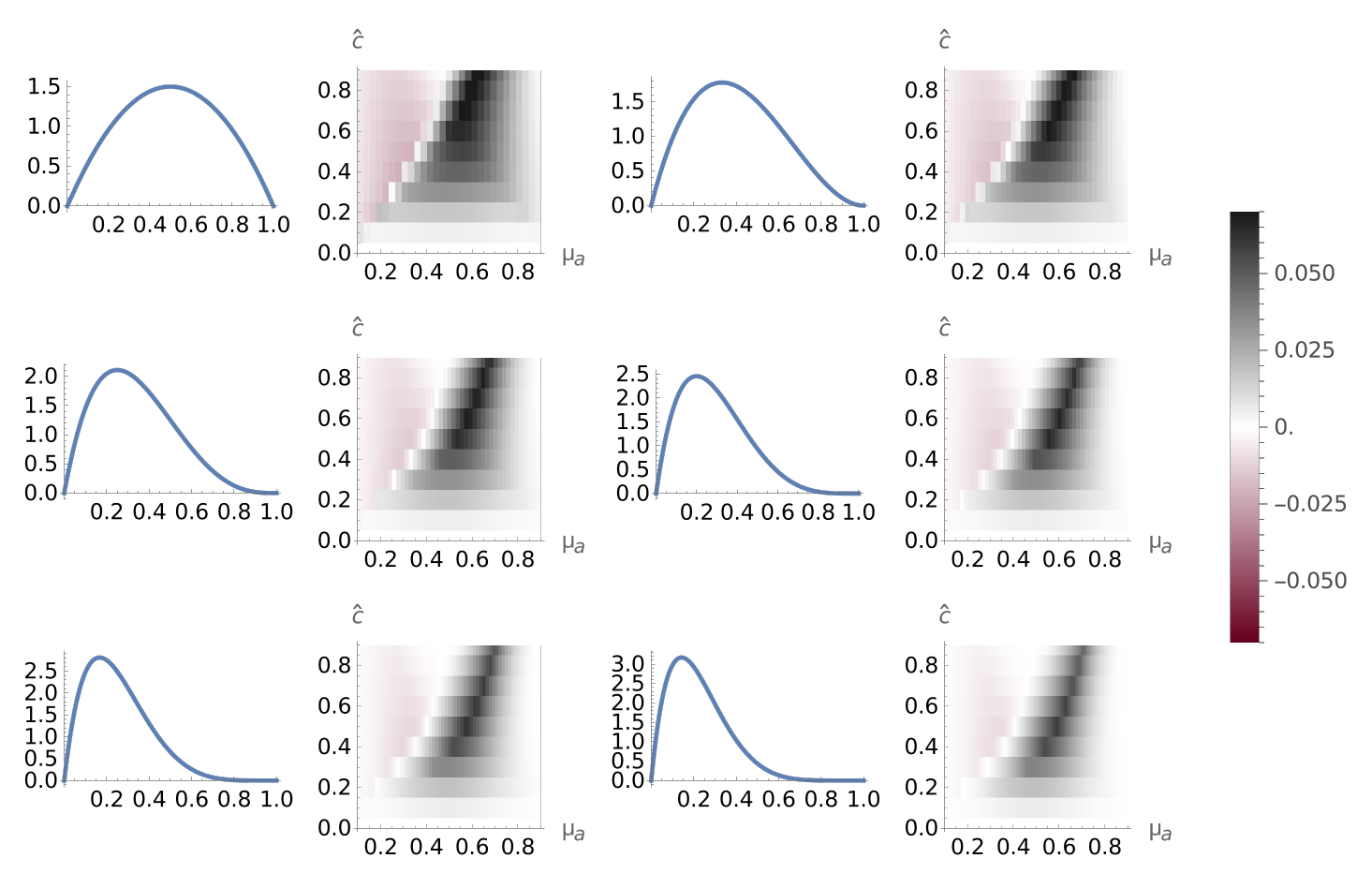}
    \caption{Junior's ex ante preference between authorship norms. Values represent the difference in Junior's expected payoff under the C-norm versus the I-norm. Positive values (gray) indicate preference for the C-norm; negative values (red) indicate preference for the I-norm; zero (white) indicates indifference between norms.}\label{f:nonbayesian-exante-Junior}
\end{figure}

\begin{figure}
    \centering
    \includegraphics[width=\textwidth]{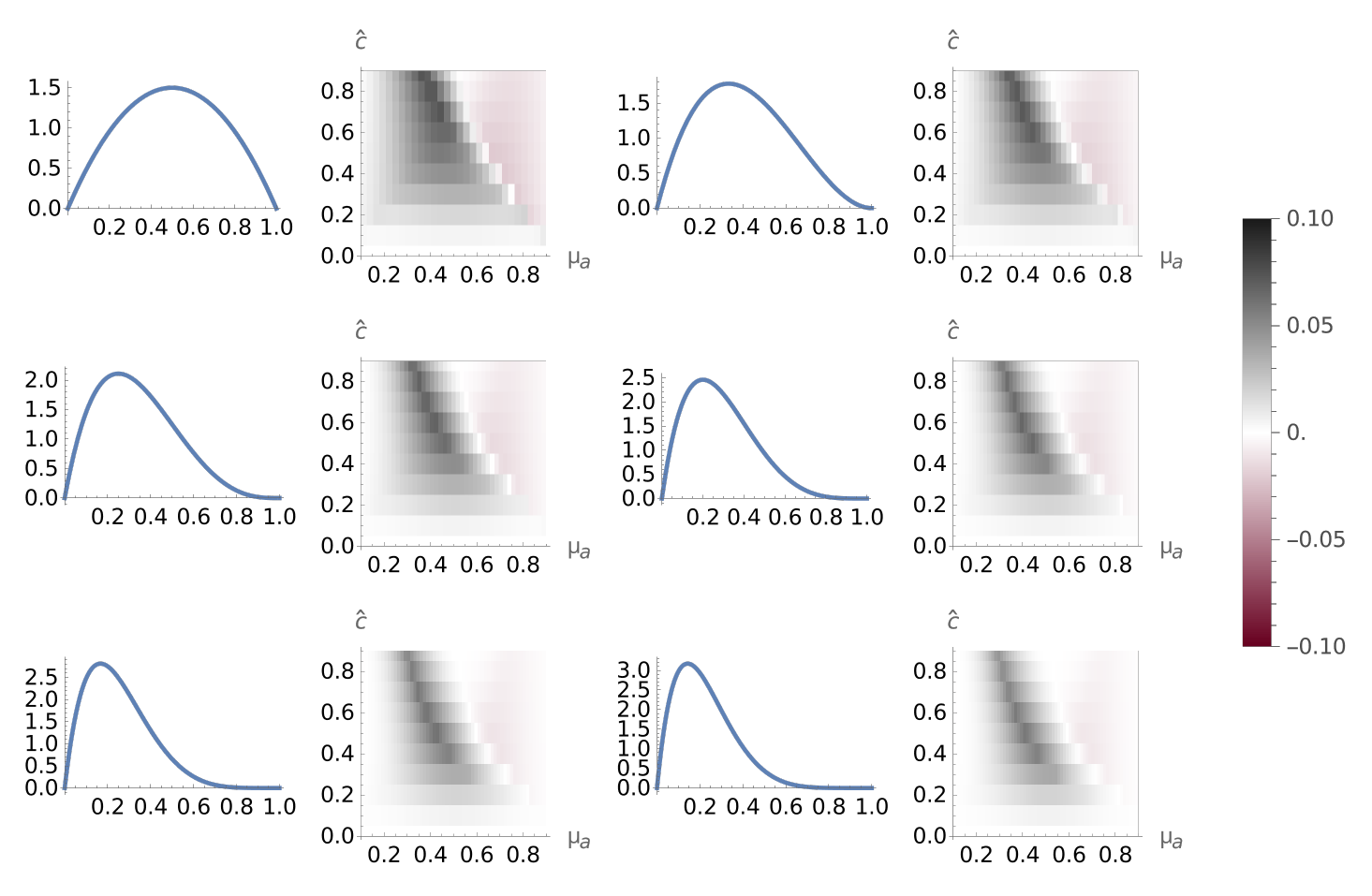}
    \caption{Senior's ex ante preference between authorship norms. Values represent the difference in Senior's expected payoff under the C-norm versus the I-norm. Positive values (gray) indicate preference for the C-norm; negative values (red) indicate preference for the I-norm; zero (white) indicates indifference between norms.}\label{f:nonbayesian-exante-Senior}
\end{figure}

We will do this from the ex ante perspective. The ex ante perspective is the perspective before they know their individual contributions.  Figures~\ref{f:nonbayesian-exante-Junior} and~\ref{f:nonbayesian-exante-Senior} illustrate the preferences between the two agents.  

There are regimes where there is agreement about the superiority of a norm and regimes where they disagree.  They tend to agree more when the expected contribution is nearly equal and $\hat{c}$ is small.  As either $\hat{c}$ becomes large or the contributions unequal, one will prefer the C-norm and the other the I-norm.

\section{Code}

Code is available at \url{https://github.com/ghostleopold/author_order}

\end{document}